\documentclass[aps,prb,twocolumn,groupedaddress]{revtex4-2}

\usepackage{multirow}
\usepackage{graphicx}

\usepackage{hyperref}
\begin{document}

\title{Leveraging Epsilon Near Zero phenomena for on-chip photonic modulation}

\author{Arun Mambra}
\email[]{itsmearun18@iisertvm.ac.in}
\author{Ravi Pant}
\author{Joy Mitra}
\email[]{j.mitra@iisertvm.ac.in}

\affiliation{School of Physics, Indian Institute of Science Education and Research Thiruvananthapuram, Kerala, India}

\date{\today}

\begin{abstract}
Epsilon-near-zero (ENZ) systems exhibit unconventional electromagnetic response close to their zero permittivity regime. Here, we explore the ability of ultrathin ENZ films to modulate the transmission of radiation from an underlying quantum emitter through active control of the carrier density of the ENZ film. The achievable on/off switching ratio is shown to be constrained by the material's loss parameter, particularly in the ENZ regime, where transmissivity increases with higher material loss. The finite loss in real materials limit the more extraordinary potential of ideal near-zero-index systems. Along with an in-depth discussion on the material parameters vis-a-vis the underlying physics, this work provides avenues to overcome the shortcomings of finite loss in real materials.  These findings are intended to guide material development and offer valuable insights for designing on-chip optical modulators and beam steering devices operating in the near-infrared regime.
\end{abstract}

\maketitle

\section{Introduction}
Metamaterials and metasurfaces have unlocked novel strategies for shaping light using plasmonic resonances, interlayer interactions, etc., with applications in imaging, sensing, cloaking, spectroscopy, and energy harvesting \cite{wang2017broadband, cortes2022optical}.
A majority of the above functionalities rely on far-field interference effects which excludes their exploitation for nano-photonic applications. The demands of nano-photonics has driven research in areas like nanoscale light sources and optical elements \cite{kalathingal2017scanning, wang2024upconversion}, and crucially in exploring novel optical environments like near zero index (NZI) media. While NZI systems have been extensively investigated in photonic crystals and metamaterials \cite{li2015chip,tang2021low,dong2021ultra,enoch2002metamaterial,ziolkowski2004propagation}, the field has been rejuvenated by the epsilon near zero (ENZ) properties of continuous media systems like degenerately doped semiconductors \cite{johns2020epsilon,johns2022tailoring,runnerstrom2018polaritonic}, conducting polymers \cite{choi2023directive,yang2019direct}, and polaritonic dielectric materials \cite{kim2016role}, opening up newer possibilities. 
ENZ systems exhibit non-trivial optical properties close to their ENZ wavelength ($\lambda_{ENZ}$), at which the real part of the material's relative permittivity ($\epsilon=\epsilon'+i\epsilon''$) goes to zero and the ENZ's optical response transitions between dielectric ($\epsilon'>0$) and metallic ($\epsilon'<0$). The ENZ regime ($|\epsilon'|<1$) exhibit properties like enhanced optical nonlinearities \cite{suresh2020enhanced, alam2016large},  super coupling \cite{silveirinha2006tunneling}, perfect absorption \cite{johns2022tailoring}, directional emission \cite{enoch2002metamaterial,choi2023directive}, slow light \cite{newman2015ferrell} and even optical levitation \cite{krasikov2014levitation}. 
The homogeneous optical properties of continuous media ENZ systems, coupled with the associated materials advantages e.g., reduced complexity of fabrication, mechanical flexibility, electrically controlled spectral tunability, homogeneous field confinement, directional emission control and more, will further expand the applications and exploration of ENZ phenomena in advancing optical and photonic technologies.
Ideal, loss-less ENZ systems \cite{liberal2017near, liberal2020near,hwang2023simultaneous} would exhibit properties like infinite phase velocity with zero group velocity \cite{liberal2016nonradiating, gong2022radiative} in their ENZ regime, which spatially maps electrodynamics to an electrostatic limit. It effectively decouples the temporal and spatial response, disentangling the electric and magnetic components \cite{liberal2017near,engheta2013pursuing,ciattoni2013polariton,javani2016real} and thus would exhibit perfect reflection and infinite impedance to wave propagation. Though the more extreme consequences are muted in real systems, due to the finite loss, the residual effects may be leveraged for manipulation of light, especially at the nanoscale.

Here, we investigate the potential of an  ultrathin ENZ film with finite loss to achieve  on/off control of transmission of radiation from an underlying quantum emitter and its spatio-temporal modulation. This is enabled by dynamically modulating the carrier density of the ENZ film via electric gating that tunes its dielectric properties, including the $\lambda_{ENZ}$. Transmission is shown to minimise when the emitter wavelength ($\lambda_{em}$) matches $\lambda_{ENZ}$, at which the ENZ film's loss parameter is, paradoxically, shown to aid transmission and thus reduce  the achievable transmission on/off ratio.
The shortcomings posed by the finite material loss may be offset by increasing the accumulation or depletion length as a fraction of the  ENZ film thickness, which is constrained by the carrier density. A straightforward multilayer design is proposed to increase the effective screening length while maintaining the flexibilities of dynamic control. Finally, a compact device architecture for on-chip beam steering  is presented, leveraging phase modulations afforded by locally gated ultrathin ENZ film.
The results provide key insights into the design and development of nanoscale optical modulators using ENZ systems, paving the way for their integration into on-chip photonic circuits.

\section{Materials and Methods}
\begin{figure}
	\includegraphics[width=7cm]{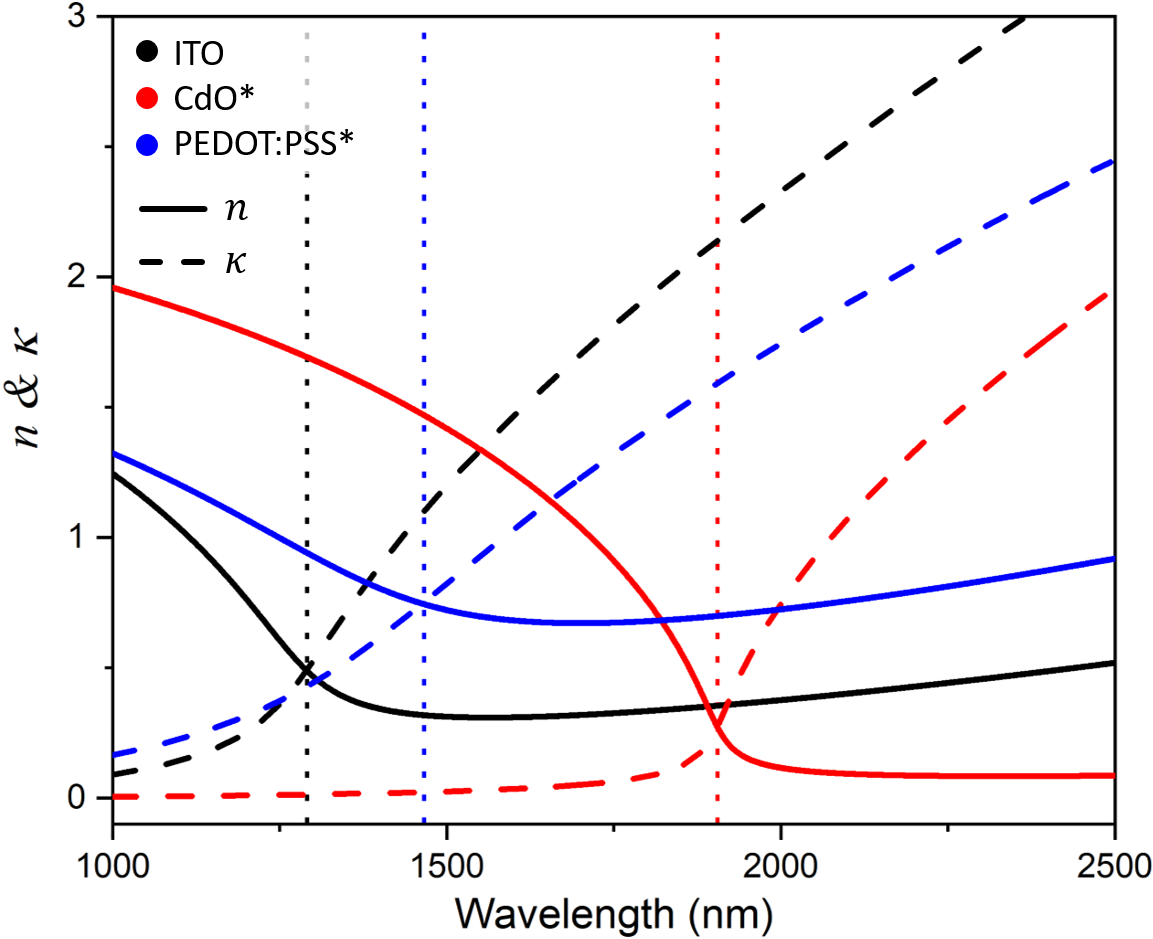} 
	\caption{Real and imaginary parts of the refractive index of ITO, CdO and PEDOT:PSS. Vertical lines denote the $\lambda_{ENZ}$, where $n=\kappa$.}
	\label{nkplot}
\end{figure}

\begin{table}
	\resizebox{0.5\textwidth}{!}{%
		\begin{tabular}{|c|c|c|c|c|}
			\hline
			& \textbf{Material} & \textbf{$\lambda_{ENZ}(\mu m)$} & \textbf{$\epsilon''$ at $\lambda_{ENZ}$} & \textbf{Ref} \\
			\hline
			\multirow{10}{*}{\rotatebox{90}{free carrier}}
			& Al & 0.083 & 0.03 & \cite{rakic1995algorithm}\\
			& Na& 0.23 & 0.003 & \cite{silvestri2024resonant}\\
			\cline{2-5}
			& ITO& 1.2 - 2.5 & 0.3 - 0.8 & \cite{johns2020epsilon}\\
			& CdO$^*$ & 1.9 & 0.14 & \cite{johns2022tailoring,sachet2015dysprosium,yang2017femtosecond}\\
			& ZnO(Al)$^*$ & 1.2 - 2 & 0.29 - 0.8 & \cite{wang2015wide,kim2016role}\\
			& Ga-ZnO & 1.19 & 0.31 & \cite{kim2016role}\\
			%& $TiN$ & \hl{0.4 - 0.6} & \hl{3.7 - 4.2} & [ref]\\
			& LABSO & 1.44 & 0.45 & \cite{kim2024perovskite}\\
			\cline{2-5}
			& PEDOT:PSS$^*$ & 0.65 - 1.7 & 0.3 - 1.2 & \cite{yang2019direct,han2024tunable}\\
			& TDBC & 0.4 & 0.16 & \cite{lee2018strong}\\
			& HTJSq & 0.52 & 0.17 & \cite{lee2018strong}\\
			\hline
			\multirow{7}{*}{\rotatebox{90}{phononic}}
			& SiO$_2$ & 8.143 & 0.46 & \cite{johns2022tailoring}\\
			& SiC & 10.3 & 0.03 & \cite{kim2016role}\\
			& AlN & 11.111 & 0.02 & \cite{passler2019second}\\
			& Si$_3$N$_4$ &	9.6 & 0.35 & \cite{li2023broadband}\\
			& Al$_2$O$_3$ & 10.6 & 0.62 & \cite{li2023broadband}\\
			& SrTiO$_3$ & 12.68, 21.05 & 0.24, 0.35 &\cite{xu2024highly}\\
			%$TiO_2$ &    &   &  \\
			\cline{2-5}
			& $h$BN & 6.2$^\parallel$, 12.35$^\perp$ & 0.08$^\parallel$, 0.05$^\perp$ & \cite{caldwell2014sub, jacob2014hyperbolic}\\
			\hline
	\end{tabular}}
	\caption{Reported ENZ materials and their  $\epsilon''$ values at $\lambda_{ENZ}$. * - denotes doping.}
	\label{tab:table1}
\end{table}
Table \ref{tab:table1} lists homogeneous materials that exhibit an ENZ regime, classified as `free carrier' and `phononic' based on the underlying origin of the materials' $\epsilon'$ going to zero. The `free carrier' basket include metals, semiconductors and conducting polymers in which the ENZ regime arises due to the collective response of their free carriers. The dielectric properties of such systems are well described by the Drude-Lorentz model, where the  free carrier density ($N_c$) determines the  $\lambda_{ENZ}$  that typically lies between the ultraviolet to the near IR. Fig. \ref{nkplot} plots the real and imaginary components of the  refractive index ($\tilde{n}=n+i\kappa$) for ITO,  doped CdO and PEDOT:PSS, with the vertical dashed lines identifying the respective $\lambda_{ENZ}$. The corresponding dielectric constants are shown in supplementary material (SM) fig. S1 \cite{SM}, in which the ENZ regimes ($|\epsilon'|<1$)  are identified with colored bands. 
The `phononic' group include oxides (SiO$_2$, Al$_2$O$_3$), nitrides (AlN) and  layered 2D materials like hBN, in which the ENZ regime originates from phonon polaritons, thus coinciding with optical phonon modes, with $\lambda_{ENZ}$ lying in the mid-IR range, as shown in table \ref{tab:table1}. Here, we focus on `free-carrier' type ENZ materials, though the electromagnetics of homogeneous ENZ systems, as discussed in section IIIA,  remains broadly applicable regardless of the origin of the ENZ response. The complementarity between free-carrier and phononic ENZ materials lies in the spectral regimes of $\lambda_{ENZ}$, their intrinsic losses, and the distinct sets of control parameters available to achieve tunability of their optical properties.

Following the Drude model, the relative permittivity of free carrier systems is given by  $\epsilon(\omega)=\epsilon_\infty-\omega^2_p/(\omega^2 +\gamma^2)+i(\omega^2_p\gamma/\omega(\omega^2+\gamma^2))$, where three material specific parameters, $\epsilon_\infty$: the high frequency relative permittivity, $\omega_p$: the plasma frequency and $\gamma$: the collision rate or damping parameter determine the system's optical response. The dependence of $\lambda_{ENZ}$ on $N_c$ as $\lambda_{ENZ}=2 \pi c / \sqrt{\omega_p^2/\epsilon_ \infty-\gamma^2 }$ provides a straightforward control over the optical properties of `free carrier' ENZ materials both via chemical  doping\cite{johns2022tailoring,johns2020epsilon} and electrostatic gating \cite{shi2016field,mambra2021dynamic,feigenbaum2010unity}, which directly controls $N_c$. 
A commercial finite element method (FEM) modeling software COMSOL Multiphysics\textsuperscript{\textregistered} 5.3a was used to conduct the simulations of light-matter interaction in ENZ media. The dipole emitter sources have been defined as current dipoles with emission wavelength ($\lambda_{em}$) = 1600 nm. The optical properties of the ENZ material were defined using the Drude model parameters, $\epsilon_\infty =$ 3.9; $\gamma =$ 10$^{10}$ Hz - 10$^{14}$ Hz and variable $N_c$ such that $\lambda_{ENZ} $ lies in the range 1340 nm – 1860 nm. The  semiconductor and wave-optics modules of the simulation package were coupled to simulate solutions to the electromagnetic wave equation due to change in dielectric properties of the ENZ system as a result of electric gating. 
Further details regarding the simulation domain and geometry of the models along with  the boundary conditions, the optimized mesh and material properties used are available at SM section S2 \cite{SM}.

\section{Results and Discussion}
\subsection{Dipole in unbounded ENZ media} \label{all_ENZ}
\begin{figure*}
	\centering
	\includegraphics[width=16.5cm]{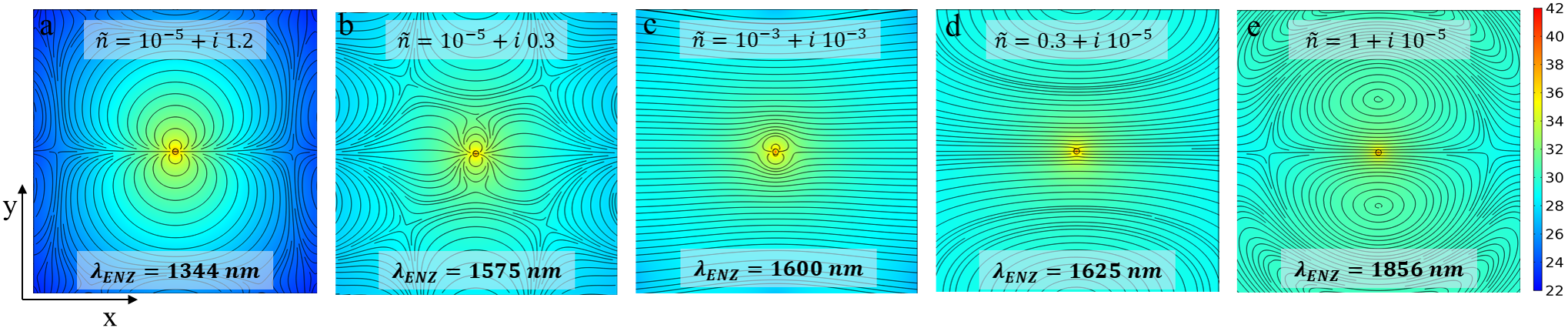}
	\caption{Electric field lines around a dipole oscillator, radiating at $\lambda_{em}$ = 1600 nm, embedded in ENZ media with (a) and (b) $\lambda_{ENZ} <  \lambda_{em}$, (c) $\lambda_{ENZ}$ = $\lambda_{em}$ and (d) and (e) $\lambda_{ENZ} > \lambda_{em}$. $\lambda_{ENZ}$ and the $\tilde{n}$ at 1600 nm are mentioned individually. Background color represents log($|\vec{E}|$)  as per the colorbar. Plot dimensions:  2.5 $\mu$m$ \times$ 2.5 $\mu$m. } 
	\label{unboundedENZ}
\end{figure*}
The physical manifestations of light–matter interactions in the ENZ regime are discussed using the example of a point dipole emitter, which emits at a free-space wavelength of $\lambda_{em}$ = 1600 nm  and angular frequency $\omega_{em}= 2\pi c/\lambda_{em}$. This emitter is encapsulated within a spherical vacuum cavity of 50 nm diameter and embedded in an unbounded ENZ medium. The ENZ media is  described by the Drude model parameters, $\gamma=10^{10}$ Hz with a variable $N_c$ and thus $\lambda_{ENZ}$. For $\lambda_{ENZ}$ varying from 1344 nm to 1856 nm the simulated electric field is presented in fig. \ref{unboundedENZ}, in which the false colored background correspond to $\log(|\vec{E}|)$ as depicted in the colorbar. The corresponding magnetic field ($\vec{H}$) plots are shown in SM fig. S5a. \cite{SM}.
At $\lambda_{em}$(=1600 nm), the ENZ medium responds like a dielectric for $\lambda_{ENZ} > \lambda_{em}$ and like a metal for $\lambda_{ENZ} < \lambda_{em}$. 
Thus for $\lambda_{ENZ}$ = 1856 nm and 1625 nm, the ENZ media is a dielectric with $\epsilon'> 0$ and radiative dipolar modes are evidenced in the $\vec{E}$ plots, figs. \ref{unboundedENZ}d and \ref{unboundedENZ}e. The values of $\tilde{n}$ for the ENZ film at 1600 nm, listed in the figures, indicate that reduction of $\lambda_{ENZ}$ towards $\lambda_{em}$ is accompanied by a decrease in $n$ at 1600 nm, which leads to spatial stretching of the $\vec{E}$ field pattern.  
For  $\lambda_{ENZ}$ $\approx$ $\lambda_{em}$, $n=\kappa \simeq 10^{-3}$ and spatially the electric field lines mimic those of a static dipole, as shown in fig. \ref{unboundedENZ}c and SM fig. S3 \cite{SM}. Notably, evolution of the $\vec{H}$ field shows a significant reduction in strength for $\lambda_{em}\approx\lambda_{ENZ}$, as shown in SM fig. S5a \cite{SM}.
For  $\lambda_{em} > \lambda_{ENZ}$, an ideal loss-less ENZ ($\epsilon'<0$ and $\epsilon'' = 0$) would respond as a lossy material with a purely imaginary $\tilde{n}$($=i\sqrt{\epsilon'}$), causing strong attenuation that prohibits propagation, but may sustain non-radiative modes  in ENZ regime\cite{gong2022radiative,liberal2016nonradiating}. However, the finite loss in real ENZ systems ensure that the $n$ remains non-zero at $\lambda_{em} > \lambda_{ENZ}$, thus allowing propagation. 
Figs. \ref{unboundedENZ}a-b show the electric field with progressively higher $N_c$, i.e., $\lambda_{ENZ}$ = 1575 and 1344 nm, that renders the ENZ system metallic at 1600 nm. Note that both cases show a small yet finite $n$ at $\lambda_{em}$, with  $\kappa$ increasing as $\lambda_{ENZ}$ decreases, resulting in stronger attenuation of the propagating modes.
\begin{figure*}
	\centering
	\includegraphics[width=12cm]{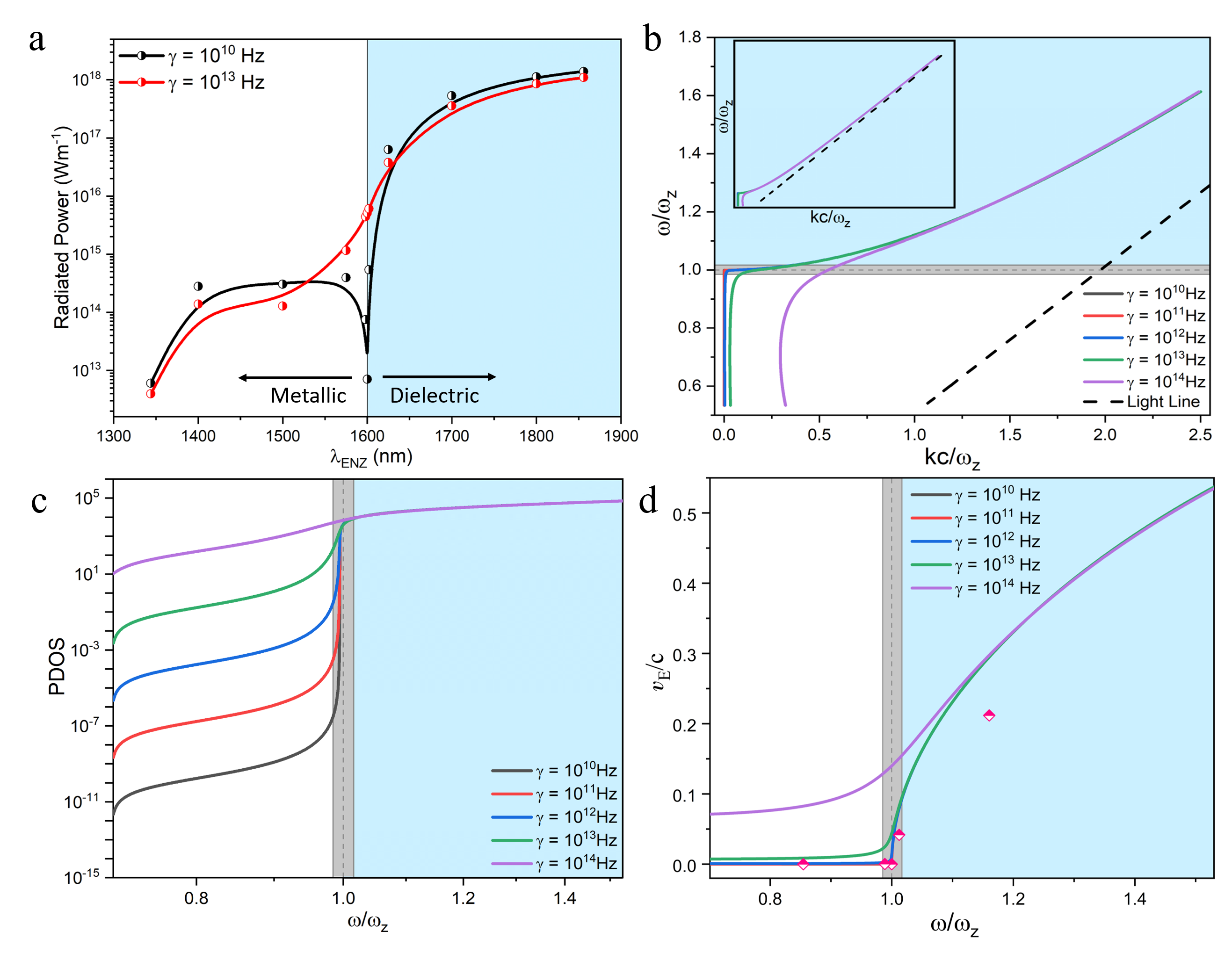}
	\caption{Properties of ENZ media for different values of the damping factor  $\gamma$. (a) Radiated power from a point dipole embedded in an unbounded ENZ media. (b) Photon dispersion relation, inset: expanded range demonstrating asymptotic convergence to the light line in the ENZ media for $\omega \rightarrow \infty$. (c) Photon density of states and (d) spectral variation of $v_E/c$, the scatter points shows simulated data and line plot shows analytically calculated data. The blue and white backgrounds demarcate dielectric and metallic regimes of the ENZ and the grey band denotes the spectral range 1600 $\pm$ 10 nm.}
	\label{PDOS-vg}
\end{figure*}
The above are reflected in the variation of power radiated across the ENZ media as its $\lambda_{ENZ}$ is changed as shown in fig. \ref{PDOS-vg}a, for two different values of $\gamma$.
For $\gamma = 10^{10}$ Hz, the radiated power is highest when the ENZ medium is a dielectric and decreases with $\lambda_{ENZ}$, reaching a minimum for $\lambda_{ENZ} = \lambda_{em}$. The radiated power increases as $\lambda_{ENZ}$ decreases below $\lambda_{em}$ but then diminishes due to stronger dissipation in the metallic medium. Spatial plots of time-averaged Poynting vector ($\langle\vec{S}\rangle$) visualizes the above in SM fig. S5b \cite{SM}.
With increased material damping ($\gamma = 10^{13}$ Hz), the radiated power decreases in both the dielectric and metallic regimes, except for $\lambda_{ENZ}\approx\lambda_{em}$, where the radiated power increases with $\gamma$ and quenches the characteristic minimum at $\lambda_{ENZ} = \lambda_{em}$.
This spectral dependence of transmission through the ENZ medium  may be understood in terms of the evolution of the optical dispersion,  photonic density of states (PDOS), and the energy velocity ($v_E$), as shown in fig. \ref{PDOS-vg}b-d. The grey bands demarcate the spectral range 1600 $\pm$ 10 nm, which is investigated in the later sections.
The dispersion plots in fig. \ref{PDOS-vg}b attest to the paucity in the allowed $\vec{k}$ values for $\omega /\omega_z \leq 1$ i.e., in the ENZ and metallic regimes, where $\omega_{z}= 2\pi c/\lambda_{ENZ}$, especially for low $\gamma$. The PDOS decreases significantly as the ENZ media transitions from a dielectric to metallic, with a distinct reduction at $\omega /\omega_z= 1$, as shown in fig. \ref{PDOS-vg}c. 
Increasing $\gamma$, increases both the  available $\vec{k}$ and the PDOS for $\omega /\omega_z\leq 1$. Thus, a non-zero $\gamma$ ensures limited yet finite $\vec{k}$ and PDOS at $\lambda_{ENZ}$, thereby allowing transmission. 
In the metallic regime of the ENZ, with  high absorption and anomalous dispersion ($dn/d\omega<0$), the group velocity $v_g$ (SM fig. S6 \cite{SM}) does not quantify the rate of energy or information transfer, which is better represented by the energy velocity.  
$v_E$ is given by the ratio of the time-averaged $\vec{S}$ to energy density ($W$), i.e. $v_E=\langle\vec{S}\rangle/\langle W\rangle$ or following Loudon \cite{loudon1970propagation}, $v_E=c/n \left(1+2\omega\kappa/n\gamma\right)^{-1}$, as depicted in fig. \ref{PDOS-vg}d  \cite{loudon1970propagation,brillouin2013wave,nunes2011electromagnetic}.
$v_E$ decreases monotonically as the ENZ medium transitions from the dielectric to metallic and goes to zero in the metallic regime, albeit for low damping. Refer to SM sec. S4 \cite{SM} for further discussions.

To summarize, around $\omega/ \omega_z\simeq 1$ coupling of light into the ENZ  medium is hampered by the limited range of allowed $\vec{k}$-vectors, diminished PDOS and an increase in electromagnetic impedance (SM fig. S4 \cite{SM}). The light that couple propagate slow as given by the diminished value of $v_E/c \sim10^{-4}$, which is commensurate with the demonstrated slow-light effects in ENZ media \cite{liberal2017near, reshef2019nonlinear, ciattoni2013polariton}.
When the emission frequency matches $\omega_z$, the spatial and temporal components of the electromagnetic field decouple in the ENZ medium, substantially reducing the propagation speed and resulting in minimizing radiated power. This spectral window with $\omega/\omega_z\simeq $1 is akin to an optical bandgap, characterized by a suppressed PDOS, low $v_E$ and elevated $Z$. 
Notably, increase in $\gamma$ increases the PDOS, broadens the range of  $\vec{k}$-vectors, increases $v_E$, while decreasing  impedance of the ENZ medium, thereby enhancing both electromagnetic coupling to and transmission.  Further, its worth noting that the width of the minimum in radiated power at $\lambda_{ENZ}$=1600 nm (fig. \ref{PDOS-vg}a) is determined by the variation of the electromagnetic impedance shown in SM fig. S4 \cite{SM}.
Absence of radiative modes in the ENZ spectral regime around $\lambda_{ENZ}$, in materials like ITO, CdO etc., and the tunability of  $\lambda_{ENZ}$ via their free carrier density can be exploited to dynamically modulate their optical properties, between high and low transmissive states. These intriguing properties of unbounded ENZ materials are retained in bounded thin-film configurations, enabling their practical applications in radiation modulation. Here, we explore device architectures that leverage the dynamic tunability of electronic ENZ thin films for on-chip signal modulation and directional beam-steering applications.

\subsection{Controlling Transmittance of an ENZ thin film}
\begin{figure}
	\centering
	\includegraphics[width=7cm]{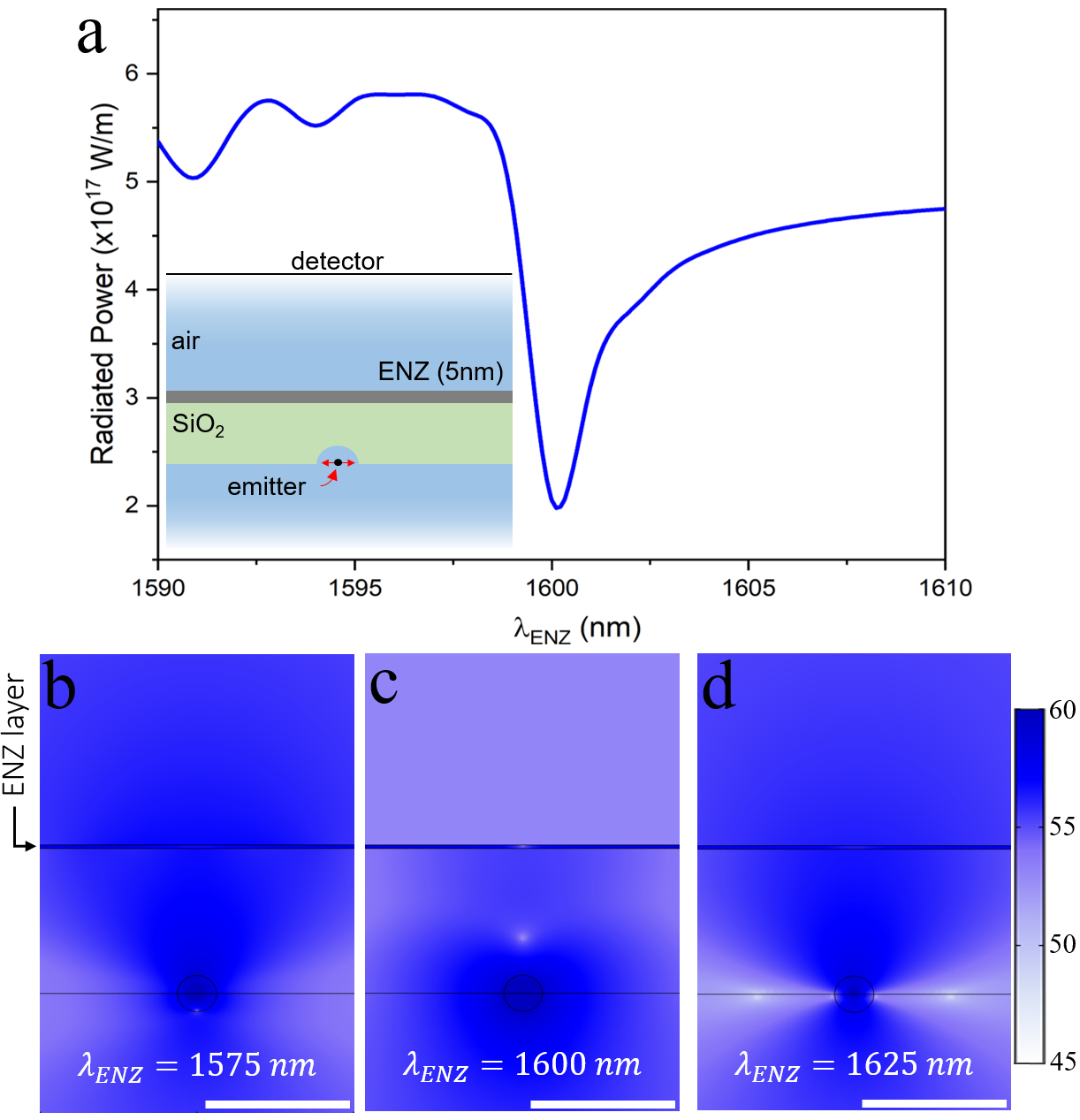}
	\caption{(a) Radiated power transmitted across a ENZ film {\it vs.} $\lambda_{ENZ}$. Inset: Schematic of the device with $5$ nm ENZ thinfilm on $SiO_2$ with a point emitter ($\lambda_{em}$=1600 nm) underneath. Magnitude of power flow $(|\vec{S}|)$  (log scale) for various $\lambda_{ENZ}$ (b) 1575 nm, (c) 1600 nm, (d) 1625 nm, scalebar: 200 nm.}
	\label{rad-power}
\end{figure}
Towards practical realization of ENZ based devices, we consider the case of an emitter {\it embedded} behind an 5 nm thick ENZ thin film on a transparent substrate (SiO$_2$) as shown in the schematic in the inset of fig. \ref{rad-power}a. The dipole oriented horizontally radiates at $\lambda_{em}$ = 1600 nm and the ENZ thinfilm is characterised by the parameters as before, with $\gamma$ = 10$^{10}$ Hz and a variable $N_c$. Power radiated by the emitter, passing through the substrate and the ENZ film is detected at the top boundary of the simulation domain and its variation  with $\lambda_{ENZ}$ is shown in fig. \ref{rad-power}a. The radiated power reaches a minima when $\lambda_{ENZ} = \lambda_{em}$ with an FWHM of 1.7 nm.
The radiation pattern around the dipole, for $\lambda_{ENZ}$ = 1575 nm, 1600 nm and 1625 nm are shown in  figs. \ref{rad-power}b-d, which plots the magnitude of the Poynting vector ($\vec{S}$) on a logarithmic scale.  In spite of the ultrathin thickness of the ENZ film, the radiation pattern is highly localized when $\lambda_{ENZ}\approx \lambda_{em}$ as shown in fig. \ref{rad-power}c, with minimum transmission. As $\lambda_{ENZ}$ shifts away from $\lambda_{em}$, transmission across the ENZ layer increases and is comparable across both the  dielectric and metallic regimes investigated here. 
The high transmission in the metallic regime of the ENZ is surprising and occurs under the action of the ENZ thinfilm's dispersion, finite loss and the film thickness being far smaller than the wavelength of radiation and the skin depth.
As discussed earlier, for free-carrier ENZ systems, the $\lambda_{ENZ}$ is tunable via its inverse proportiality to  $ \sqrt{N_c}$ \cite{johns2020epsilon,huang2016gate} and in such systems $N_c \simeq 10^{25} - 10^{27}$/m$^3$.
The model  in the inset of fig. \ref{rad-power}a is modified to include a 1 nm HfO$_2$ insulating layer  and top gate  over the ENZ film to allow modulation of $N_c$, and thus $\lambda_{ENZ}$, by the application of a gate bias (V$_G$). 
\begin{figure}
	\centering
	\includegraphics[width=6cm]{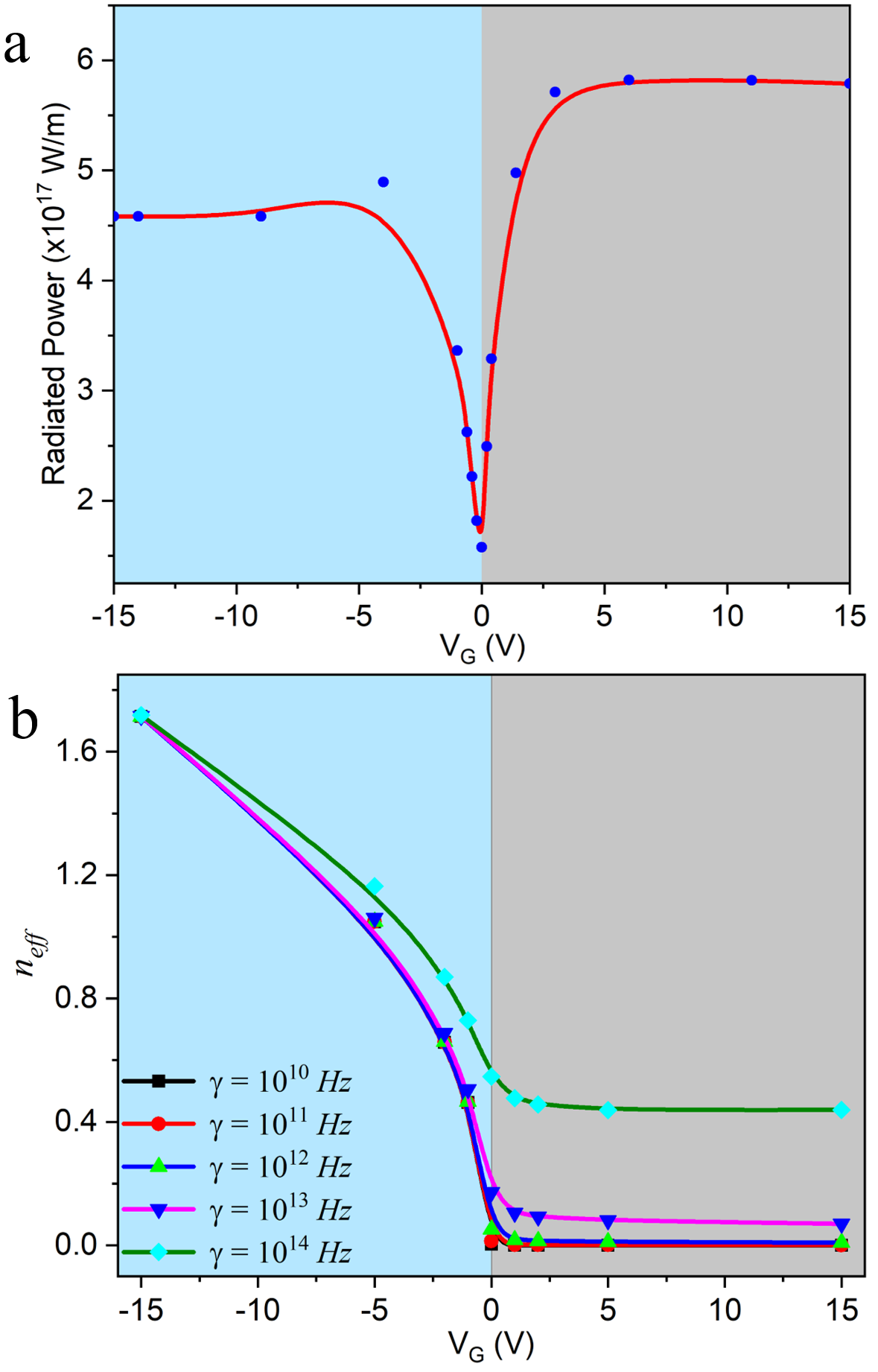}
	\caption{$V_G$ dependence of (a) power transmitted through an ENZ film ($\gamma = 10^{10}$ Hz) and (b) effective refractive index ($n_{eff}$) of the ENZ film with different $\gamma$, for $\lambda_{em}$=1600 nm.}
	\label{neff-vg}
\end{figure}
Variation in power transmitted across an ENZ thinfilm with V$_G$ is shown in fig. \ref{neff-vg}a, which is minimum at V$_G$ = 0 V due to the non-radiative modes of the ENZ. The carrier density of the film ($N_c = 6 \times 10^{26}$/m$^3$) is chosen such that its  $\lambda_{ENZ} $ =  1600 nm, matching the $\lambda_{em}$ of the emitter at zero gate bias.
Under positive (negative) V$_G$, accumulation (depletion) of electrons in the ENZ thin film results in a blue (red) shift of the thinfilm's effective $\lambda_{ENZ}$, making its response metallic (dielectric) at $\lambda_{em}$ = 1600 nm. 
Variation of V$_G$ between $\pm$15 V results in spatially variable $N_c(w)$ within the ENZ film, with average $\langle N_c\rangle$  of 1.13$\times$10$^{26}$ m$^{-3}$ and 1.68$\times$10$^{27}$ m$^{-3}$ and depletion and accumulation widths of ($w\simeq$) 3 nm and 1 nm, as shown in SM fig. S7 \cite{SM}.
The resultant spatially varying $\tilde{n}(w)$, the real and imaginary parts of which are shown in SM fig. S9 \cite{SM}, allows V$_G$ dependent control of an effective refractive index $n_{eff}$ at 1600 nm, averaged across the cross-section of the ENZ thinfilm, as depicted in fig. \ref{neff-vg}b.
The similarity of the $n_{eff}-V_G$ plot with that of $n$ {\it vs.} $\omega$, as depicted in fig. S8a of \cite{SM}, is noteworthy, suggesting that varying V$_G$ is equivalent to probing the different optical regimes of the ENZ film at a fixed emitter wavelength.
Increase in radiated power for negative V$_G$, i.e., depletion, is expected, since the ENZ film behaves as a dielectric at 1600 nm, with its reduced $N_c$ and thus a red-shifted effective $\lambda_{ENZ}$. Under positive V$_G$ the electron accumulation layer blue-shifts the effective $\lambda_{ENZ}$ and the ENZ behaves like a metal at 1600 nm. 
For any given $\gamma$, the higher $\kappa$ in the metallic regime has little effect due to the ultra thin nature of the film, as discussed before. 
Thus, V$_G$ effectively controls transmissivity of the ENZ layer demonstrating a straightforward approach for  modulating the emission intensity from an embedded emitter. Though a 1 nm HfO$_2$ layer has been employed above to maximize gate control, a 5 nm HfO$_2$ layer also yields comparable transmission modulation albeit with a higher $V_G$ (SM Sec. S8  \cite{SM}). While the intensity modulation on-off ratio for the 5 nm film for $\gamma=10^{10}$ Hz is around 66\% (-4.77 dB) as shown in fig. \ref{neff-vg}a, the modulation frequency will be determined by the temporal response of the gate, i.e. carrier mobility in the ENZ layer.
Note that the effective on/off ratio of transmission i.e. intensity modulation via spectral tuning of $\lambda_{ENZ}$ with $\lambda_{em}$ is determined by the damping factor ($\gamma$) of the ENZ medium and its thickness. In the ENZ regime, particularly when $\epsilon'\approx0$, the imaginary component ($\epsilon''$) significantly contributes to $\tilde{n}$ and thus controls transmission through the ENZ medium. Thus, crucial to achieving the full potential of ENZ effect is the development of ENZ materials with small $\gamma$. SM fig. S10a \cite{SM} illustrates variation in power transmitted through an ENZ film as the $\gamma$ decreases from $10^{13} - 10^7$ Hz, reinforcing the adverse effect of increasing $\gamma$, which effectively introduces radiative modes near $\lambda_{ENZ}$. This is better understood in the context of the point emitter embedded in an unbounded ENZ media (SM Sec. S9) \cite{SM}. SM figs. S12a-e \cite{SM} shows the evolution of an iso-power surface of constant value ($3 \times 10^{21}$ W/m$^2$), for increasing $\gamma$ with $\lambda_{ENZ}=\lambda_{em}$. The background  plots $|\vec{S}|$ in log scale and shows that  as $\gamma$ increases the iso-power surface enlarges, denoting higher propagation. 
Thus, as real part of the dielectric constant and the refractive index becomes smaller in the ENZ regime, the thickness of the ENZ thinfilm i.e., the optical path length becomes increasingly crucial in determining the transmitted power. SM fig. S10b \cite{SM} shows the variation in radiated power for ENZ film thicknesses ranging from 5 nm to 105 nm. SM fig. S10c,d \cite{SM} plots the radiated power on/off ratio P(1590)/P(1600) as functions of ln($\gamma$) and film thickness, highlighting the ability to modulate transmissive radiation in ENZ media by adjusting its carrier density and thickness. 

\begin{figure}
	\includegraphics[width=\columnwidth]{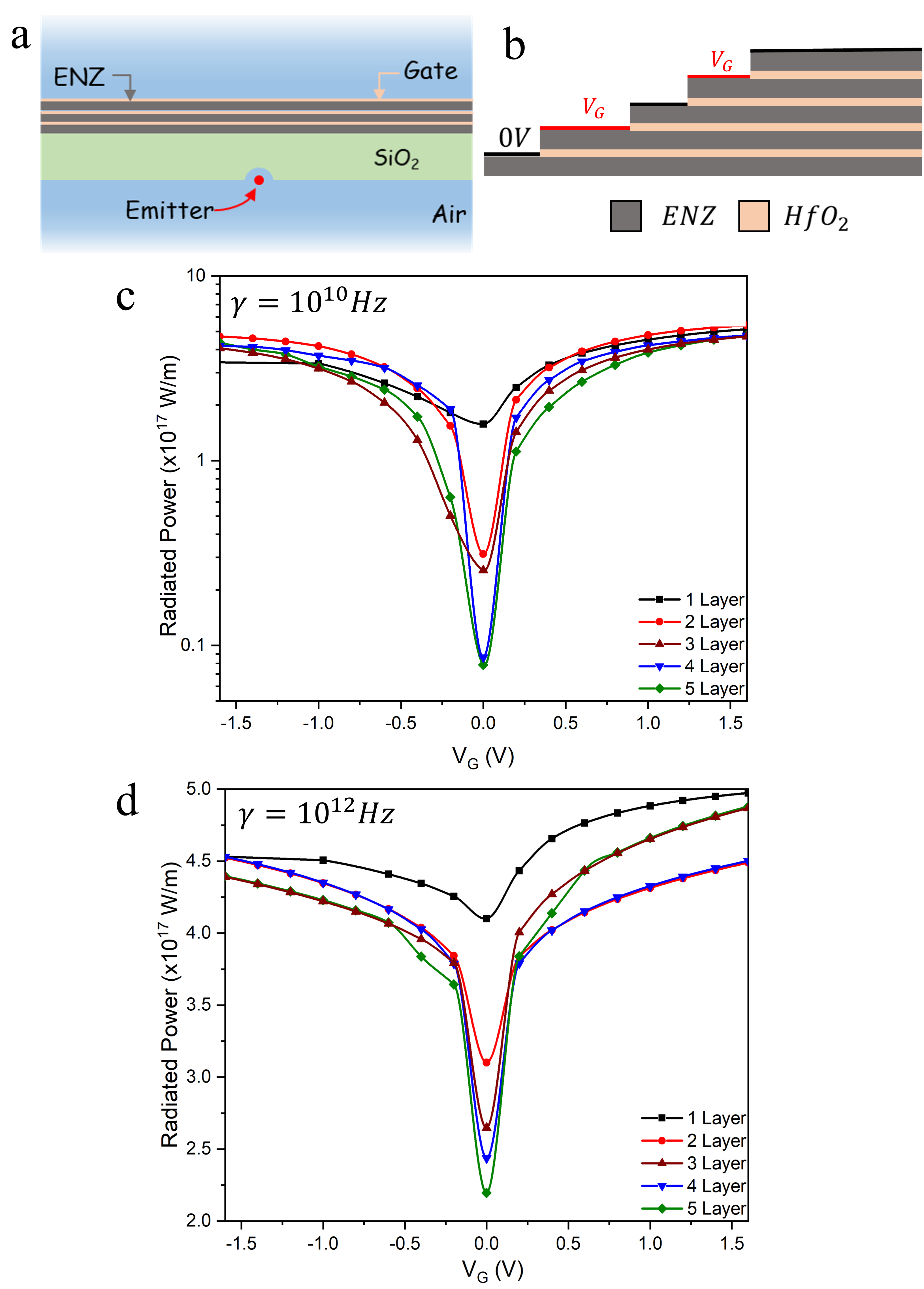}
	\caption{(a) Multilayer ENZ device geometry for emission tailoring. (b) Zoomed up schematic of the multilayer gated system for dynamic control.
		Radiated power versus V$_G$ for different number of layers with (c) $\gamma=10^{10}$ Hz and (d) $\gamma=10^{12}$ Hz. The $\lambda_{em}=1600$ nm and the non-gated $\lambda_{ENZ}=1600$ nm.}
	\label{multilayer-scheme}
\end{figure}	
At $\lambda_{ENZ}$, free carrier ENZ materials have $\epsilon''$ ranging from $0.001$ to $0.6$ that corresponds to $\gamma$ between $10^{12} -10^{14} Hz$ (table \ref{tab:table1}), hence exhibit weak radiation attentuation, especially for ultra-thin films with thickness $\sim$ 5 nm.  While thicker films would show higher transmitted power on/off ratio, the efficacy of gating controlled tuning of $\lambda_{ENZ}$ is limited, as screening of the gate field restricts the depletion/accumulation widths, typically around 2 nm from the interface \cite{mambra2021dynamic, maniyara2019tunable}, thereby negating change in the $\tilde{n}_{eff}$. The problem can be mitigated by employing a multi-layer ENZ thin-film design, interleaved with an insulating dielectric, as shown in fig. \ref{multilayer-scheme}.
Here, multiple 5 nm thick layers of ENZ films are separated  by a 1 nm thick dielectric gate layer (HfO$_2)$, where each ENZ film can be gated with respect to the adjoining ENZ film. This multi-layer scheme negates the requirement of individual transparent gate electrodes, where alternate ENZ films are biased at 0V and V$_G$, allowing each ENZ film to either  accumulate or deplete carriers, as shown in fig. \ref{multilayer-scheme}.
SM fig. S14a \cite{SM} shows the spatial variation of electron density across a five layer stack of ENZ films with $N_c = 6 \times 10^{26}$/m$^3$ and  $\lambda_{ENZ} $ =  1600 nm. For V$_G$ varying between $\pm$ 1.5 V, electrons accumulate and deplete in the alternating ENZ layers rendering them metallic or dielectric at 1600 nm. Since, transmission across the individual layers increase irrespective of the change towards metallic or dielectric, overall the stack becomes more transmissive for both positive and negative V$_G$.   
The efficacy of the multilayer gated structure is evidenced in the radiated power plots shown in fig. \ref{multilayer-scheme}c-d for 1 - 5 layers of ENZ film with  $\gamma$ = 10$^{10}$ Hz and 10$^{12}$ Hz, demonstrating transmission modulation up to 98\% (-18dB) and 50\% (-3dB), respectively. SM fig. S15 shows the case of increased $\gamma$ (10$^{13}$ Hz), with 1-7 layers \cite{SM}. This architecture provides improved response  compared to a single layer  utilizing a multilayer gating scheme that removes the requirement of separate gate electrodes. 

\subsection{Beam steering}
\begin{figure*}[t]
	\centering
	\includegraphics[width=16cm]{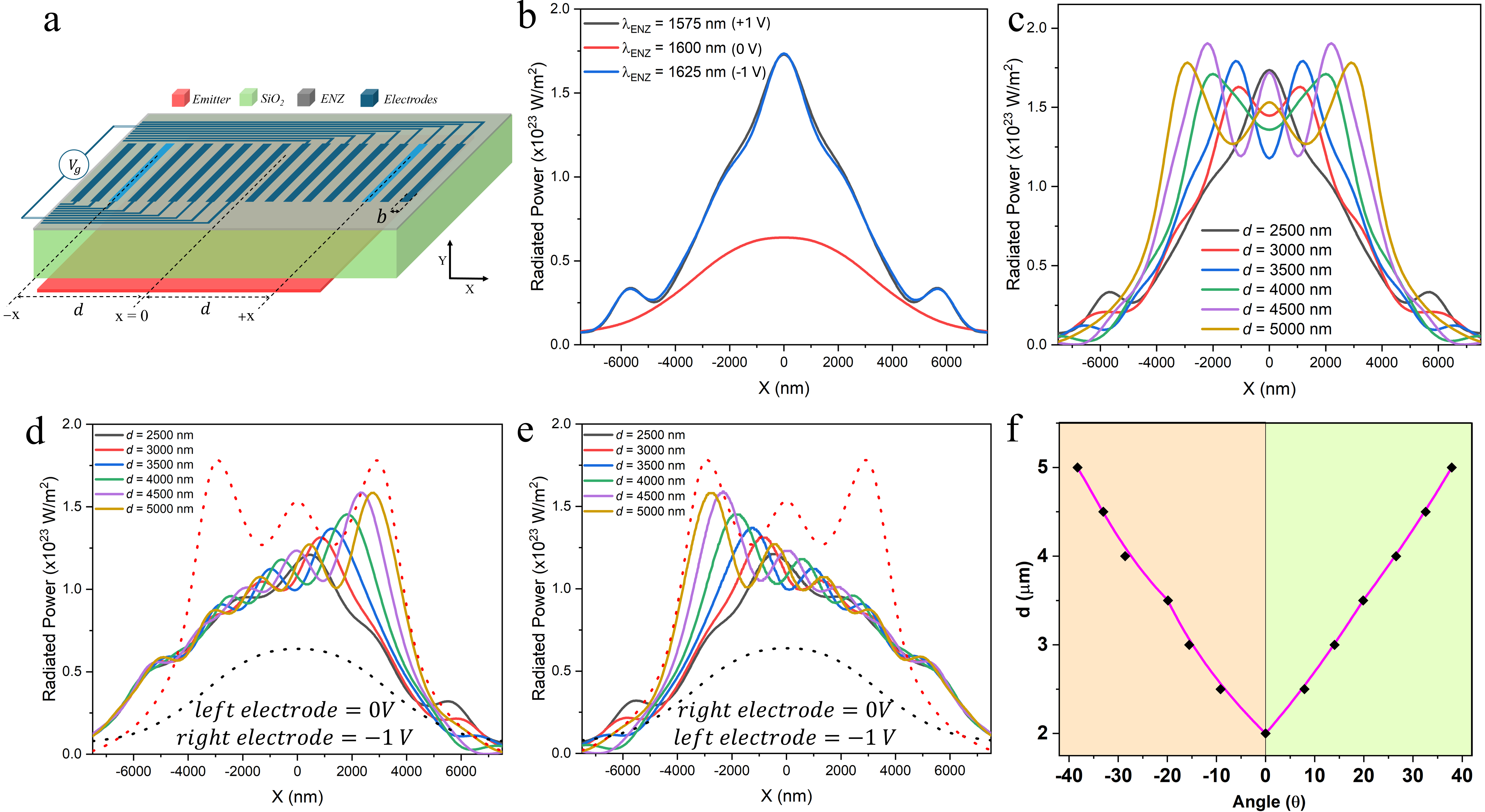}
	\caption{(a) Schematic for effective beam steering device; non-zero V$_G$ applied to a symmetric pair of electrodes (light blue) separated by $2d$.  (b) Lateral variation in radiated intensity 3.6 $\mu$m above the device for different V$_G$ applied to the active gate electrodes spaced at $d$ = 2.5 $\mu$m. Spatial variation in the radiated intensity for varying $d$, with the transmitting windows at (c) $\lambda_{ENZ}=1625$ nm, (d) electrodes on the left (-x) at 0 V ($\lambda_{ENZ}=1600$ nm) and electrodes on the right (+x) at -1 V ($\lambda_{ENZ}=1625$) and (e) conversely electrode on left (-x) at -1V ($\lambda_{ENZ}=1625$ nm) and right (+x) at 0V ($\lambda_{ENZ}=1600$ nm), b = 100 nm and $\lambda_{em}=1600$ nm (f) Quantifying beam steering in terms of angle of deviation in degrees.}
	\label{beam-steering}
\end{figure*}

Leveraging the intensity modulation enabled by gating the ENZ thin film, we propose the design of an on-chip beam steering device. Here, a single 5 nm thick ENZ thin film with $\lambda_{ENZ}$ of 1600 nm is coated on a transparent substrate. The underside of the substrate features an extended array of dipole emitters, extending laterally over 10 $\mu$m. As depicted in Figure \ref{beam-steering}a, here the gate electrodes form a liner array of interdigitated fingers with a width of 100 nm and a separation of 100 nm, extending along the x-direction over 10 $\mu$m. A gate voltage, V$_G$ is applied to a symmetric pair of gate electrodes (shown in light blue, fig. \ref{beam-steering}a), equidistant from the centre and separated by a distance of $2d$, with $d$ varying from 100 nm to 5000 nm.
Since the $\lambda_{ENZ}$ of the film matches the emission wavelength of the embedded dipoles, transmission across the ENZ film is minimum. However, once a non-zero V$_G$ is applied at the pair of gates, the transmittance of the ENZ film below the gate electrodes is modified locally. Note that the change in transmittance is independent of the sign of V$_G$, as shown in fig. \ref{neff-vg}a and fig. \ref{multilayer-scheme}. Fig. \ref{beam-steering}b plots the lateral variation in power detected by an extended detector placed along the top edge of the simulation domain (3.6 $\mu$m above ENZ layer), corresponding to V$_G$ = 0, and V$_G=\pm$ 1V for which the effective $\lambda_{ENZ}$ of the gated windows are 1600 nm (red line), 1575 nm (black line) and 1625 nm (blue line). The V$_G$ = 0 case denotes the lowest transmission state, which increases locally as the windows are rendered dielectric or metallic by changing $\lambda_{ENZ}$.
The scheme mimics Young’s double-slit experimental setup, with the exception that the entire ENZ film has non-zero transmission, which is highest at the regions with non-zero V$_G$. Consequently, the detected intensity profile at the top edge shows signatures of constructive and destructive interference, as shown in fig. \ref{beam-steering}c, for $d$ varying from 2.5 - 5 $\mu$m. The device typically offers a $\sim$2.5 times change in radiated intensity at the detectors while switching between the ``on'' state (red line) and the ``off'' state (black or blue line) in fig. \ref{beam-steering}b, equivalent to 60\% to 75\% (-4 to -6 db) intensity modulation. Comprehending the interference pattern is complex due to the non-trivial phase shift incurred by light at the interface of ultrathin ENZ films with Re($\tilde{n}$) below one, and has been discussed elsewhere \cite{johns2022tailoring}. The beam-steering capability is demonstrated by simulating the lateral variation of intensity at the detector array by applying V$_G$ = -1V to only one of the pair of electrodes (discussed above), for various $d$ across the electrode array, as shown in figs. \ref{beam-steering}d-e. The black dotted lines correspond to zero bias applied at all gate electrodes and the red dotted lines represent the condition where two symmetric electrodes (separated by 2$d = 10 \mu$m) are biased at -1 V. The highest intensity point on the detector shifts laterally as the position ($d$) of the single ``on’’ electrode shifts outwards towards the left (-x) or right (+x), and it is accompanied by an increase in peak intensity. Results in figs. \ref{beam-steering}d-e show that the peak intensity detected at a distance 3.6 $\mu$m from the ENZ film laterally shifts by $\sim \pm 3 \mu$m as the ``on'' electrode is shifted by $\pm 5 \mu$m from the centre. This is equivalent to a beam steering capability of $\pm 38^\circ $, as shown in fig. \ref{beam-steering}f. Interference across the light transmitted through the entire ENZ thin film determines the intensity pattern at the detector array and also quantifies the transmission ``on-off" ratio, which is a crucial parameter for achieving emission control and beam steering in the context of on-chip photonic devices. The design is readily scalable and can accommodate wider sources resulting in higher angular steering capability, upto $\pm 47^\circ$, as shown in SM fig. S16 \cite{SM}. Such scalable architecture can be readily integrated into on-chip photonic circuits, enabling applications in optical communications, data routing and free-space optical interconnects.

\section{Conclusion}
This study provides key insights into light-matter interactions in homogeneous ENZ materials, elucidating the challenges posed by material damping in harnessing the benefits of the ENZ regime for functional device design and engineering. It demonstrates strategies for on-chip modulation of radiation intensity within the near-infrared spectral range. A straight-forward device design is investigated, where dynamic tuning of free-carrier density in 5 nm thick ENZ thinfilms via gating enables modulation of transmissivity as a function of gate voltage. The study emphasises the significance of carrier density (screening length) and material losses in defining the efficiency (on/off ratio) of ENZ-based intensity modulators. To address these limitations, a multilayer architecture is proposed that enhances the effective optical interaction length while maintaining dynamic tunability, offering a scalable solution for improving the performance of practical ENZ-based photonic devices. Finally, dynamic beam steering is demonstrated utilising spatially localised phase modulation enabled by selective control of ultrathin ENZ films. Future advancements in low-loss ENZ materials development and improved gating architectures will further enhance the efficiency and applicability of these strategies for integration in next-generation on-chip optical and photonic technologies.

\begin{acknowledgements}
	The authors acknowledge ISTEM, Government of India, for access to the COMSOL Multiphysics software and financial support from SERB, Government of India (No.CRG/2023/006878). AM acknowledges PhD fellowship from IISER Thiruvananthapuram.
\end{acknowledgements}

\bibliography{Bibfile}

\begin{thebibliography}{53}%
\makeatletter
\providecommand \@ifxundefined [1]{%
 \@ifx{#1\undefined}
}%
\providecommand \@ifnum [1]{%
 \ifnum #1\expandafter \@firstoftwo
 \else \expandafter \@secondoftwo
 \fi
}%
\providecommand \@ifx [1]{%
 \ifx #1\expandafter \@firstoftwo
 \else \expandafter \@secondoftwo
 \fi
}%
\providecommand \natexlab [1]{#1}%
\providecommand \enquote  [1]{``#1''}%
\providecommand \bibnamefont  [1]{#1}%
\providecommand \bibfnamefont [1]{#1}%
\providecommand \citenamefont [1]{#1}%
\providecommand \href@noop [0]{\@secondoftwo}%
\providecommand \href [0]{\begingroup \@sanitize@url \@href}%
\providecommand \@href[1]{\@@startlink{#1}\@@href}%
\providecommand \@@href[1]{\endgroup#1\@@endlink}%
\providecommand \@sanitize@url [0]{\catcode `\\12\catcode `\$12\catcode
  `\&12\catcode `\#12\catcode `\^12\catcode `\_12\catcode `\%12\relax}%
\providecommand \@@startlink[1]{}%
\providecommand \@@endlink[0]{}%
\providecommand \url  [0]{\begingroup\@sanitize@url \@url }%
\providecommand \@url [1]{\endgroup\@href {#1}{\urlprefix }}%
\providecommand \urlprefix  [0]{URL }%
\providecommand \Eprint [0]{\href }%
\providecommand \doibase [0]{https://doi.org/}%
\providecommand \selectlanguage [0]{\@gobble}%
\providecommand \bibinfo  [0]{\@secondoftwo}%
\providecommand \bibfield  [0]{\@secondoftwo}%
\providecommand \translation [1]{[#1]}%
\providecommand \BibitemOpen [0]{}%
\providecommand \bibitemStop [0]{}%
\providecommand \bibitemNoStop [0]{.\EOS\space}%
\providecommand \EOS [0]{\spacefactor3000\relax}%
\providecommand \BibitemShut  [1]{\csname bibitem#1\endcsname}%
\let\auto@bib@innerbib\@empty
%</preamble>
\bibitem [{\citenamefont {Wang}\ \emph {et~al.}(2017)\citenamefont {Wang},
  \citenamefont {Wu}, \citenamefont {Su}, \citenamefont {Lai}, \citenamefont
  {Hung~Chu}, \citenamefont {Chen}, \citenamefont {Lu}, \citenamefont {Chen},
  \citenamefont {Xu}, \citenamefont {Kuan} \emph {et~al.}}]{wang2017broadband}%
  \BibitemOpen
  \bibfield  {author} {\bibinfo {author} {\bibfnamefont {S.}~\bibnamefont
  {Wang}}, \bibinfo {author} {\bibfnamefont {P.~C.}\ \bibnamefont {Wu}},
  \bibinfo {author} {\bibfnamefont {V.-C.}\ \bibnamefont {Su}}, \bibinfo
  {author} {\bibfnamefont {Y.-C.}\ \bibnamefont {Lai}}, \bibinfo {author}
  {\bibfnamefont {C.}~\bibnamefont {Hung~Chu}}, \bibinfo {author}
  {\bibfnamefont {J.-W.}\ \bibnamefont {Chen}}, \bibinfo {author}
  {\bibfnamefont {S.-H.}\ \bibnamefont {Lu}}, \bibinfo {author} {\bibfnamefont
  {J.}~\bibnamefont {Chen}}, \bibinfo {author} {\bibfnamefont {B.}~\bibnamefont
  {Xu}}, \bibinfo {author} {\bibfnamefont {C.-H.}\ \bibnamefont {Kuan}}, \emph
  {et~al.},\ }\bibfield  {title} {\bibinfo {title} {Broadband achromatic
  optical metasurface devices},\ }\href@noop {} {\bibfield  {journal} {\bibinfo
   {journal} {Nature communications}\ }\textbf {\bibinfo {volume} {8}},\
  \bibinfo {pages} {187} (\bibinfo {year} {2017})}\BibitemShut {NoStop}%
\bibitem [{\citenamefont {Cort{\'e}s}\ \emph {et~al.}(2022)\citenamefont
  {Cort{\'e}s}, \citenamefont {Wendisch}, \citenamefont {Sortino},
  \citenamefont {Mancini}, \citenamefont {Ezendam}, \citenamefont {Saris},
  \citenamefont {de~S.~Menezes}, \citenamefont {Tittl}, \citenamefont {Ren},\
  and\ \citenamefont {Maier}}]{cortes2022optical}%
  \BibitemOpen
  \bibfield  {author} {\bibinfo {author} {\bibfnamefont {E.}~\bibnamefont
  {Cort{\'e}s}}, \bibinfo {author} {\bibfnamefont {F.~J.}\ \bibnamefont
  {Wendisch}}, \bibinfo {author} {\bibfnamefont {L.}~\bibnamefont {Sortino}},
  \bibinfo {author} {\bibfnamefont {A.}~\bibnamefont {Mancini}}, \bibinfo
  {author} {\bibfnamefont {S.}~\bibnamefont {Ezendam}}, \bibinfo {author}
  {\bibfnamefont {S.}~\bibnamefont {Saris}}, \bibinfo {author} {\bibfnamefont
  {L.}~\bibnamefont {de~S.~Menezes}}, \bibinfo {author} {\bibfnamefont
  {A.}~\bibnamefont {Tittl}}, \bibinfo {author} {\bibfnamefont
  {H.}~\bibnamefont {Ren}},\ and\ \bibinfo {author} {\bibfnamefont {S.~A.}\
  \bibnamefont {Maier}},\ }\bibfield  {title} {\bibinfo {title} {Optical
  metasurfaces for energy conversion},\ }\href@noop {} {\bibfield  {journal}
  {\bibinfo  {journal} {Chemical reviews}\ }\textbf {\bibinfo {volume} {122}},\
  \bibinfo {pages} {15082} (\bibinfo {year} {2022})}\BibitemShut {NoStop}%
\bibitem [{\citenamefont {Kalathingal}\ \emph {et~al.}(2017)\citenamefont
  {Kalathingal}, \citenamefont {Dawson},\ and\ \citenamefont
  {Mitra}}]{kalathingal2017scanning}%
  \BibitemOpen
  \bibfield  {author} {\bibinfo {author} {\bibfnamefont {V.}~\bibnamefont
  {Kalathingal}}, \bibinfo {author} {\bibfnamefont {P.}~\bibnamefont
  {Dawson}},\ and\ \bibinfo {author} {\bibfnamefont {J.}~\bibnamefont
  {Mitra}},\ }\bibfield  {title} {\bibinfo {title} {Scanning tunnelling
  microscope light emission: Finite temperature current noise and over cut-off
  emission},\ }\href@noop {} {\bibfield  {journal} {\bibinfo  {journal}
  {Scientific reports}\ }\textbf {\bibinfo {volume} {7}},\ \bibinfo {pages}
  {3530} (\bibinfo {year} {2017})}\BibitemShut {NoStop}%
\bibitem [{\citenamefont {Wang}\ \emph {et~al.}(2024)\citenamefont {Wang},
  \citenamefont {Kalathingal}, \citenamefont {Trushin}, \citenamefont {Liu},
  \citenamefont {Wang}, \citenamefont {Guo}, \citenamefont {{\"O}zyilmaz},
  \citenamefont {Nijhuis},\ and\ \citenamefont {Eda}}]{wang2024upconversion}%
  \BibitemOpen
  \bibfield  {author} {\bibinfo {author} {\bibfnamefont {Z.}~\bibnamefont
  {Wang}}, \bibinfo {author} {\bibfnamefont {V.}~\bibnamefont {Kalathingal}},
  \bibinfo {author} {\bibfnamefont {M.}~\bibnamefont {Trushin}}, \bibinfo
  {author} {\bibfnamefont {J.}~\bibnamefont {Liu}}, \bibinfo {author}
  {\bibfnamefont {J.}~\bibnamefont {Wang}}, \bibinfo {author} {\bibfnamefont
  {Y.}~\bibnamefont {Guo}}, \bibinfo {author} {\bibfnamefont {B.}~\bibnamefont
  {{\"O}zyilmaz}}, \bibinfo {author} {\bibfnamefont {C.~A.}\ \bibnamefont
  {Nijhuis}},\ and\ \bibinfo {author} {\bibfnamefont {G.}~\bibnamefont {Eda}},\
  }\bibfield  {title} {\bibinfo {title} {Upconversion electroluminescence in 2d
  semiconductors integrated with plasmonic tunnel junctions},\ }\href@noop {}
  {\bibfield  {journal} {\bibinfo  {journal} {Nature Nanotechnology}\ ,\
  \bibinfo {pages} {1}} (\bibinfo {year} {2024})}\BibitemShut {NoStop}%
\bibitem [{\citenamefont {Li}\ \emph {et~al.}(2015)\citenamefont {Li},
  \citenamefont {Kita}, \citenamefont {Mu{\~n}oz}, \citenamefont {Reshef},
  \citenamefont {Vulis}, \citenamefont {Yin}, \citenamefont {Lon{\v{c}}ar},\
  and\ \citenamefont {Mazur}}]{li2015chip}%
  \BibitemOpen
  \bibfield  {author} {\bibinfo {author} {\bibfnamefont {Y.}~\bibnamefont
  {Li}}, \bibinfo {author} {\bibfnamefont {S.}~\bibnamefont {Kita}}, \bibinfo
  {author} {\bibfnamefont {P.}~\bibnamefont {Mu{\~n}oz}}, \bibinfo {author}
  {\bibfnamefont {O.}~\bibnamefont {Reshef}}, \bibinfo {author} {\bibfnamefont
  {D.~I.}\ \bibnamefont {Vulis}}, \bibinfo {author} {\bibfnamefont
  {M.}~\bibnamefont {Yin}}, \bibinfo {author} {\bibfnamefont {M.}~\bibnamefont
  {Lon{\v{c}}ar}},\ and\ \bibinfo {author} {\bibfnamefont {E.}~\bibnamefont
  {Mazur}},\ }\bibfield  {title} {\bibinfo {title} {On-chip zero-index
  metamaterials},\ }\href@noop {} {\bibfield  {journal} {\bibinfo  {journal}
  {Nature Photonics}\ }\textbf {\bibinfo {volume} {9}},\ \bibinfo {pages} {738}
  (\bibinfo {year} {2015})}\BibitemShut {NoStop}%
\bibitem [{\citenamefont {Tang}\ \emph {et~al.}(2021)\citenamefont {Tang},
  \citenamefont {DeVault}, \citenamefont {Camayd-Mu{\~n}oz}, \citenamefont
  {Liu}, \citenamefont {Jia}, \citenamefont {Du}, \citenamefont {Mello},
  \citenamefont {Vulis}, \citenamefont {Li},\ and\ \citenamefont
  {Mazur}}]{tang2021low}%
  \BibitemOpen
  \bibfield  {author} {\bibinfo {author} {\bibfnamefont {H.}~\bibnamefont
  {Tang}}, \bibinfo {author} {\bibfnamefont {C.}~\bibnamefont {DeVault}},
  \bibinfo {author} {\bibfnamefont {S.~A.}\ \bibnamefont {Camayd-Mu{\~n}oz}},
  \bibinfo {author} {\bibfnamefont {Y.}~\bibnamefont {Liu}}, \bibinfo {author}
  {\bibfnamefont {D.}~\bibnamefont {Jia}}, \bibinfo {author} {\bibfnamefont
  {F.}~\bibnamefont {Du}}, \bibinfo {author} {\bibfnamefont {O.}~\bibnamefont
  {Mello}}, \bibinfo {author} {\bibfnamefont {D.~I.}\ \bibnamefont {Vulis}},
  \bibinfo {author} {\bibfnamefont {Y.}~\bibnamefont {Li}},\ and\ \bibinfo
  {author} {\bibfnamefont {E.}~\bibnamefont {Mazur}},\ }\bibfield  {title}
  {\bibinfo {title} {Low-loss zero-index materials},\ }\href@noop {} {\bibfield
   {journal} {\bibinfo  {journal} {Nano letters}\ }\textbf {\bibinfo {volume}
  {21}},\ \bibinfo {pages} {914} (\bibinfo {year} {2021})}\BibitemShut
  {NoStop}%
\bibitem [{\citenamefont {Dong}\ \emph {et~al.}(2021)\citenamefont {Dong},
  \citenamefont {Liang}, \citenamefont {Camayd-Mu{\~n}oz}, \citenamefont {Liu},
  \citenamefont {Tang}, \citenamefont {Kita}, \citenamefont {Chen},
  \citenamefont {Wu}, \citenamefont {Chu}, \citenamefont {Mazur} \emph
  {et~al.}}]{dong2021ultra}%
  \BibitemOpen
  \bibfield  {author} {\bibinfo {author} {\bibfnamefont {T.}~\bibnamefont
  {Dong}}, \bibinfo {author} {\bibfnamefont {J.}~\bibnamefont {Liang}},
  \bibinfo {author} {\bibfnamefont {S.}~\bibnamefont {Camayd-Mu{\~n}oz}},
  \bibinfo {author} {\bibfnamefont {Y.}~\bibnamefont {Liu}}, \bibinfo {author}
  {\bibfnamefont {H.}~\bibnamefont {Tang}}, \bibinfo {author} {\bibfnamefont
  {S.}~\bibnamefont {Kita}}, \bibinfo {author} {\bibfnamefont {P.}~\bibnamefont
  {Chen}}, \bibinfo {author} {\bibfnamefont {X.}~\bibnamefont {Wu}}, \bibinfo
  {author} {\bibfnamefont {W.}~\bibnamefont {Chu}}, \bibinfo {author}
  {\bibfnamefont {E.}~\bibnamefont {Mazur}}, \emph {et~al.},\ }\bibfield
  {title} {\bibinfo {title} {Ultra-low-loss on-chip zero-index materials},\
  }\href@noop {} {\bibfield  {journal} {\bibinfo  {journal} {Light: Science \&
  Applications}\ }\textbf {\bibinfo {volume} {10}},\ \bibinfo {pages} {10}
  (\bibinfo {year} {2021})}\BibitemShut {NoStop}%
\bibitem [{\citenamefont {Enoch}\ \emph {et~al.}(2002)\citenamefont {Enoch},
  \citenamefont {Tayeb}, \citenamefont {Sabouroux}, \citenamefont
  {Gu{\'e}rin},\ and\ \citenamefont {Vincent}}]{enoch2002metamaterial}%
  \BibitemOpen
  \bibfield  {author} {\bibinfo {author} {\bibfnamefont {S.}~\bibnamefont
  {Enoch}}, \bibinfo {author} {\bibfnamefont {G.}~\bibnamefont {Tayeb}},
  \bibinfo {author} {\bibfnamefont {P.}~\bibnamefont {Sabouroux}}, \bibinfo
  {author} {\bibfnamefont {N.}~\bibnamefont {Gu{\'e}rin}},\ and\ \bibinfo
  {author} {\bibfnamefont {P.}~\bibnamefont {Vincent}},\ }\bibfield  {title}
  {\bibinfo {title} {A metamaterial for directive emission},\ }\href@noop {}
  {\bibfield  {journal} {\bibinfo  {journal} {Physical review letters}\
  }\textbf {\bibinfo {volume} {89}},\ \bibinfo {pages} {213902} (\bibinfo
  {year} {2002})}\BibitemShut {NoStop}%
\bibitem [{\citenamefont {Ziolkowski}(2004)}]{ziolkowski2004propagation}%
  \BibitemOpen
  \bibfield  {author} {\bibinfo {author} {\bibfnamefont {R.~W.}\ \bibnamefont
  {Ziolkowski}},\ }\bibfield  {title} {\bibinfo {title} {Propagation in and
  scattering from a matched metamaterial having a zero index of refraction},\
  }\href@noop {} {\bibfield  {journal} {\bibinfo  {journal} {Physical Review
  E}\ }\textbf {\bibinfo {volume} {70}},\ \bibinfo {pages} {046608} (\bibinfo
  {year} {2004})}\BibitemShut {NoStop}%
\bibitem [{\citenamefont {Johns}\ \emph {et~al.}(2020)\citenamefont {Johns},
  \citenamefont {Puthoor}, \citenamefont {Gopalakrishnan}, \citenamefont
  {Mishra}, \citenamefont {Pant},\ and\ \citenamefont
  {Mitra}}]{johns2020epsilon}%
  \BibitemOpen
  \bibfield  {author} {\bibinfo {author} {\bibfnamefont {B.}~\bibnamefont
  {Johns}}, \bibinfo {author} {\bibfnamefont {N.~M.}\ \bibnamefont {Puthoor}},
  \bibinfo {author} {\bibfnamefont {H.}~\bibnamefont {Gopalakrishnan}},
  \bibinfo {author} {\bibfnamefont {A.}~\bibnamefont {Mishra}}, \bibinfo
  {author} {\bibfnamefont {R.}~\bibnamefont {Pant}},\ and\ \bibinfo {author}
  {\bibfnamefont {J.}~\bibnamefont {Mitra}},\ }\bibfield  {title} {\bibinfo
  {title} {Epsilon-near-zero response in indium tin oxide thin films: Octave
  span tuning and ir plasmonics},\ }\href@noop {} {\bibfield  {journal}
  {\bibinfo  {journal} {Journal of Applied Physics}\ }\textbf {\bibinfo
  {volume} {127}} (\bibinfo {year} {2020})}\BibitemShut {NoStop}%
\bibitem [{\citenamefont {Johns}\ \emph {et~al.}(2022)\citenamefont {Johns},
  \citenamefont {Chattopadhyay},\ and\ \citenamefont
  {Mitra}}]{johns2022tailoring}%
  \BibitemOpen
  \bibfield  {author} {\bibinfo {author} {\bibfnamefont {B.}~\bibnamefont
  {Johns}}, \bibinfo {author} {\bibfnamefont {S.}~\bibnamefont
  {Chattopadhyay}},\ and\ \bibinfo {author} {\bibfnamefont {J.}~\bibnamefont
  {Mitra}},\ }\bibfield  {title} {\bibinfo {title} {Tailoring infrared
  absorption and thermal emission with ultrathin film interferences in
  epsilon-near-zero media},\ }\href@noop {} {\bibfield  {journal} {\bibinfo
  {journal} {Advanced Photonics Research}\ }\textbf {\bibinfo {volume} {3}},\
  \bibinfo {pages} {2100153} (\bibinfo {year} {2022})}\BibitemShut {NoStop}%
\bibitem [{\citenamefont {Runnerstrom}\ \emph {et~al.}(2018)\citenamefont
  {Runnerstrom}, \citenamefont {Kelley}, \citenamefont {Folland}, \citenamefont
  {Nolen}, \citenamefont {Engheta}, \citenamefont {Caldwell},\ and\
  \citenamefont {Maria}}]{runnerstrom2018polaritonic}%
  \BibitemOpen
  \bibfield  {author} {\bibinfo {author} {\bibfnamefont {E.~L.}\ \bibnamefont
  {Runnerstrom}}, \bibinfo {author} {\bibfnamefont {K.~P.}\ \bibnamefont
  {Kelley}}, \bibinfo {author} {\bibfnamefont {T.~G.}\ \bibnamefont {Folland}},
  \bibinfo {author} {\bibfnamefont {J.~R.}\ \bibnamefont {Nolen}}, \bibinfo
  {author} {\bibfnamefont {N.}~\bibnamefont {Engheta}}, \bibinfo {author}
  {\bibfnamefont {J.~D.}\ \bibnamefont {Caldwell}},\ and\ \bibinfo {author}
  {\bibfnamefont {J.-P.}\ \bibnamefont {Maria}},\ }\bibfield  {title} {\bibinfo
  {title} {Polaritonic hybrid-epsilon-near-zero modes: beating the plasmonic
  confinement vs propagation-length trade-off with doped cadmium oxide
  bilayers},\ }\href@noop {} {\bibfield  {journal} {\bibinfo  {journal} {Nano
  letters}\ }\textbf {\bibinfo {volume} {19}},\ \bibinfo {pages} {948}
  (\bibinfo {year} {2018})}\BibitemShut {NoStop}%
\bibitem [{\citenamefont {Choi}\ \emph {et~al.}(2023)\citenamefont {Choi},
  \citenamefont {Kim}, \citenamefont {Wu}, \citenamefont {D’al{\'e}o},\ and\
  \citenamefont {Lee}}]{choi2023directive}%
  \BibitemOpen
  \bibfield  {author} {\bibinfo {author} {\bibfnamefont {K.-R.}\ \bibnamefont
  {Choi}}, \bibinfo {author} {\bibfnamefont {M.}~\bibnamefont {Kim}}, \bibinfo
  {author} {\bibfnamefont {J.~W.}\ \bibnamefont {Wu}}, \bibinfo {author}
  {\bibfnamefont {A.}~\bibnamefont {D’al{\'e}o}},\ and\ \bibinfo {author}
  {\bibfnamefont {Y.~U.}\ \bibnamefont {Lee}},\ }\bibfield  {title} {\bibinfo
  {title} {Directive emission from polymeric fluorophore with epsilon-near-zero
  squaraine molecular film},\ }\href@noop {} {\bibfield  {journal} {\bibinfo
  {journal} {Nanophotonics}\ }\textbf {\bibinfo {volume} {12}},\ \bibinfo
  {pages} {2471} (\bibinfo {year} {2023})}\BibitemShut {NoStop}%
\bibitem [{\citenamefont {Yang}\ \emph {et~al.}(2019)\citenamefont {Yang},
  \citenamefont {Almossalami}, \citenamefont {Wang}, \citenamefont {Wu},
  \citenamefont {Wang}, \citenamefont {Sun}, \citenamefont {Yang},\ and\
  \citenamefont {Ye}}]{yang2019direct}%
  \BibitemOpen
  \bibfield  {author} {\bibinfo {author} {\bibfnamefont {J.}~\bibnamefont
  {Yang}}, \bibinfo {author} {\bibfnamefont {H.~A.}\ \bibnamefont
  {Almossalami}}, \bibinfo {author} {\bibfnamefont {Z.}~\bibnamefont {Wang}},
  \bibinfo {author} {\bibfnamefont {K.}~\bibnamefont {Wu}}, \bibinfo {author}
  {\bibfnamefont {C.}~\bibnamefont {Wang}}, \bibinfo {author} {\bibfnamefont
  {K.}~\bibnamefont {Sun}}, \bibinfo {author} {\bibfnamefont {Y.~M.}\
  \bibnamefont {Yang}},\ and\ \bibinfo {author} {\bibfnamefont
  {H.}~\bibnamefont {Ye}},\ }\bibfield  {title} {\bibinfo {title} {Direct
  observations of surface plasmon polaritons in highly conductive organic thin
  film},\ }\href@noop {} {\bibfield  {journal} {\bibinfo  {journal} {ACS
  applied materials \& interfaces}\ }\textbf {\bibinfo {volume} {11}},\
  \bibinfo {pages} {39132} (\bibinfo {year} {2019})}\BibitemShut {NoStop}%
\bibitem [{\citenamefont {Kim}\ \emph {et~al.}(2016)\citenamefont {Kim},
  \citenamefont {Dutta}, \citenamefont {Naik}, \citenamefont {Giles},
  \citenamefont {Bezares}, \citenamefont {Ellis}, \citenamefont {Tischler},
  \citenamefont {Mahmoud}, \citenamefont {Caglayan}, \citenamefont {Glembocki}
  \emph {et~al.}}]{kim2016role}%
  \BibitemOpen
  \bibfield  {author} {\bibinfo {author} {\bibfnamefont {J.}~\bibnamefont
  {Kim}}, \bibinfo {author} {\bibfnamefont {A.}~\bibnamefont {Dutta}}, \bibinfo
  {author} {\bibfnamefont {G.~V.}\ \bibnamefont {Naik}}, \bibinfo {author}
  {\bibfnamefont {A.~J.}\ \bibnamefont {Giles}}, \bibinfo {author}
  {\bibfnamefont {F.~J.}\ \bibnamefont {Bezares}}, \bibinfo {author}
  {\bibfnamefont {C.~T.}\ \bibnamefont {Ellis}}, \bibinfo {author}
  {\bibfnamefont {J.~G.}\ \bibnamefont {Tischler}}, \bibinfo {author}
  {\bibfnamefont {A.~M.}\ \bibnamefont {Mahmoud}}, \bibinfo {author}
  {\bibfnamefont {H.}~\bibnamefont {Caglayan}}, \bibinfo {author}
  {\bibfnamefont {O.~J.}\ \bibnamefont {Glembocki}}, \emph {et~al.},\
  }\bibfield  {title} {\bibinfo {title} {Role of epsilon-near-zero substrates
  in the optical response of plasmonic antennas},\ }\href@noop {} {\bibfield
  {journal} {\bibinfo  {journal} {Optica}\ }\textbf {\bibinfo {volume} {3}},\
  \bibinfo {pages} {339} (\bibinfo {year} {2016})}\BibitemShut {NoStop}%
\bibitem [{\citenamefont {Suresh}\ \emph {et~al.}(2020)\citenamefont {Suresh},
  \citenamefont {Reshef}, \citenamefont {Alam}, \citenamefont {Upham},
  \citenamefont {Karimi},\ and\ \citenamefont {Boyd}}]{suresh2020enhanced}%
  \BibitemOpen
  \bibfield  {author} {\bibinfo {author} {\bibfnamefont {S.}~\bibnamefont
  {Suresh}}, \bibinfo {author} {\bibfnamefont {O.}~\bibnamefont {Reshef}},
  \bibinfo {author} {\bibfnamefont {M.~Z.}\ \bibnamefont {Alam}}, \bibinfo
  {author} {\bibfnamefont {J.}~\bibnamefont {Upham}}, \bibinfo {author}
  {\bibfnamefont {M.}~\bibnamefont {Karimi}},\ and\ \bibinfo {author}
  {\bibfnamefont {R.~W.}\ \bibnamefont {Boyd}},\ }\bibfield  {title} {\bibinfo
  {title} {Enhanced nonlinear optical responses of layered epsilon-near-zero
  metamaterials at visible frequencies},\ }\href@noop {} {\bibfield  {journal}
  {\bibinfo  {journal} {Acs Photonics}\ }\textbf {\bibinfo {volume} {8}},\
  \bibinfo {pages} {125} (\bibinfo {year} {2020})}\BibitemShut {NoStop}%
\bibitem [{\citenamefont {Alam}\ \emph {et~al.}(2016)\citenamefont {Alam},
  \citenamefont {De~Leon},\ and\ \citenamefont {Boyd}}]{alam2016large}%
  \BibitemOpen
  \bibfield  {author} {\bibinfo {author} {\bibfnamefont {M.~Z.}\ \bibnamefont
  {Alam}}, \bibinfo {author} {\bibfnamefont {I.}~\bibnamefont {De~Leon}},\ and\
  \bibinfo {author} {\bibfnamefont {R.~W.}\ \bibnamefont {Boyd}},\ }\bibfield
  {title} {\bibinfo {title} {Large optical nonlinearity of indium tin oxide in
  its epsilon-near-zero region},\ }\href@noop {} {\bibfield  {journal}
  {\bibinfo  {journal} {Science}\ }\textbf {\bibinfo {volume} {352}},\ \bibinfo
  {pages} {795} (\bibinfo {year} {2016})}\BibitemShut {NoStop}%
\bibitem [{\citenamefont {Silveirinha}\ and\ \citenamefont
  {Engheta}(2006)}]{silveirinha2006tunneling}%
  \BibitemOpen
  \bibfield  {author} {\bibinfo {author} {\bibfnamefont {M.}~\bibnamefont
  {Silveirinha}}\ and\ \bibinfo {author} {\bibfnamefont {N.}~\bibnamefont
  {Engheta}},\ }\bibfield  {title} {\bibinfo {title} {Tunneling of
  electromagnetic energy through subwavelength channels and bends using
  epsilon-near-zero materials},\ }\href@noop {} {\bibfield  {journal} {\bibinfo
   {journal} {Physical review letters}\ }\textbf {\bibinfo {volume} {97}},\
  \bibinfo {pages} {157403} (\bibinfo {year} {2006})}\BibitemShut {NoStop}%
\bibitem [{\citenamefont {Newman}\ \emph {et~al.}(2015)\citenamefont {Newman},
  \citenamefont {Cortes}, \citenamefont {Atkinson}, \citenamefont {Pramanik},
  \citenamefont {DeCorby},\ and\ \citenamefont {Jacob}}]{newman2015ferrell}%
  \BibitemOpen
  \bibfield  {author} {\bibinfo {author} {\bibfnamefont {W.~D.}\ \bibnamefont
  {Newman}}, \bibinfo {author} {\bibfnamefont {C.~L.}\ \bibnamefont {Cortes}},
  \bibinfo {author} {\bibfnamefont {J.}~\bibnamefont {Atkinson}}, \bibinfo
  {author} {\bibfnamefont {S.}~\bibnamefont {Pramanik}}, \bibinfo {author}
  {\bibfnamefont {R.~G.}\ \bibnamefont {DeCorby}},\ and\ \bibinfo {author}
  {\bibfnamefont {Z.}~\bibnamefont {Jacob}},\ }\bibfield  {title} {\bibinfo
  {title} {Ferrell--berreman modes in plasmonic epsilon-near-zero media},\
  }\href@noop {} {\bibfield  {journal} {\bibinfo  {journal} {Acs Photonics}\
  }\textbf {\bibinfo {volume} {2}},\ \bibinfo {pages} {2} (\bibinfo {year}
  {2015})}\BibitemShut {NoStop}%
\bibitem [{\citenamefont {Krasikov}\ \emph {et~al.}(2014)\citenamefont
  {Krasikov}, \citenamefont {Iorsh}, \citenamefont {Shalin},\ and\
  \citenamefont {Belov}}]{krasikov2014levitation}%
  \BibitemOpen
  \bibfield  {author} {\bibinfo {author} {\bibfnamefont {S.}~\bibnamefont
  {Krasikov}}, \bibinfo {author} {\bibfnamefont {I.~V.}\ \bibnamefont {Iorsh}},
  \bibinfo {author} {\bibfnamefont {A.}~\bibnamefont {Shalin}},\ and\ \bibinfo
  {author} {\bibfnamefont {P.~A.}\ \bibnamefont {Belov}},\ }\bibfield  {title}
  {\bibinfo {title} {Levitation of finite-size electric dipole over
  epsilon-near-zero metamaterial},\ }\href@noop {} {\bibfield  {journal}
  {\bibinfo  {journal} {physica status solidi (RRL)--Rapid Research Letters}\
  }\textbf {\bibinfo {volume} {8}},\ \bibinfo {pages} {1015} (\bibinfo {year}
  {2014})}\BibitemShut {NoStop}%
\bibitem [{\citenamefont {Liberal}\ and\ \citenamefont
  {Engheta}(2017)}]{liberal2017near}%
  \BibitemOpen
  \bibfield  {author} {\bibinfo {author} {\bibfnamefont {I.}~\bibnamefont
  {Liberal}}\ and\ \bibinfo {author} {\bibfnamefont {N.}~\bibnamefont
  {Engheta}},\ }\bibfield  {title} {\bibinfo {title} {Near-zero refractive
  index photonics},\ }\href@noop {} {\bibfield  {journal} {\bibinfo  {journal}
  {Nature Photonics}\ }\textbf {\bibinfo {volume} {11}},\ \bibinfo {pages}
  {149} (\bibinfo {year} {2017})}\BibitemShut {NoStop}%
\bibitem [{\citenamefont {Liberal}\ \emph {et~al.}(2020)\citenamefont
  {Liberal}, \citenamefont {Lobet}, \citenamefont {Li},\ and\ \citenamefont
  {Engheta}}]{liberal2020near}%
  \BibitemOpen
  \bibfield  {author} {\bibinfo {author} {\bibfnamefont {I.}~\bibnamefont
  {Liberal}}, \bibinfo {author} {\bibfnamefont {M.}~\bibnamefont {Lobet}},
  \bibinfo {author} {\bibfnamefont {Y.}~\bibnamefont {Li}},\ and\ \bibinfo
  {author} {\bibfnamefont {N.}~\bibnamefont {Engheta}},\ }\bibfield  {title}
  {\bibinfo {title} {Near-zero-index media as electromagnetic ideal fluids},\
  }\href@noop {} {\bibfield  {journal} {\bibinfo  {journal} {Proceedings of the
  National Academy of Sciences}\ }\textbf {\bibinfo {volume} {117}},\ \bibinfo
  {pages} {24050} (\bibinfo {year} {2020})}\BibitemShut {NoStop}%
\bibitem [{\citenamefont {Hwang}\ \emph {et~al.}(2023)\citenamefont {Hwang},
  \citenamefont {Xu},\ and\ \citenamefont {Raman}}]{hwang2023simultaneous}%
  \BibitemOpen
  \bibfield  {author} {\bibinfo {author} {\bibfnamefont {J.~S.}\ \bibnamefont
  {Hwang}}, \bibinfo {author} {\bibfnamefont {J.}~\bibnamefont {Xu}},\ and\
  \bibinfo {author} {\bibfnamefont {A.~P.}\ \bibnamefont {Raman}},\ }\bibfield
  {title} {\bibinfo {title} {Simultaneous control of spectral and directional
  emissivity with gradient epsilon-near-zero inas photonic structures},\
  }\href@noop {} {\bibfield  {journal} {\bibinfo  {journal} {Advanced
  Materials}\ }\textbf {\bibinfo {volume} {35}},\ \bibinfo {pages} {2302956}
  (\bibinfo {year} {2023})}\BibitemShut {NoStop}%
\bibitem [{\citenamefont {Liberal}\ and\ \citenamefont
  {Engheta}(2016)}]{liberal2016nonradiating}%
  \BibitemOpen
  \bibfield  {author} {\bibinfo {author} {\bibfnamefont {I.}~\bibnamefont
  {Liberal}}\ and\ \bibinfo {author} {\bibfnamefont {N.}~\bibnamefont
  {Engheta}},\ }\bibfield  {title} {\bibinfo {title} {Nonradiating and
  radiating modes excited by quantum emitters in open epsilon-near-zero
  cavities},\ }\href@noop {} {\bibfield  {journal} {\bibinfo  {journal}
  {Science advances}\ }\textbf {\bibinfo {volume} {2}},\ \bibinfo {pages}
  {e1600987} (\bibinfo {year} {2016})}\BibitemShut {NoStop}%
\bibitem [{\citenamefont {Gong}\ \emph {et~al.}(2022)\citenamefont {Gong},
  \citenamefont {Liberal}, \citenamefont {Camacho}, \citenamefont {Spreng},
  \citenamefont {Engheta},\ and\ \citenamefont {Munday}}]{gong2022radiative}%
  \BibitemOpen
  \bibfield  {author} {\bibinfo {author} {\bibfnamefont {T.}~\bibnamefont
  {Gong}}, \bibinfo {author} {\bibfnamefont {I.}~\bibnamefont {Liberal}},
  \bibinfo {author} {\bibfnamefont {M.}~\bibnamefont {Camacho}}, \bibinfo
  {author} {\bibfnamefont {B.}~\bibnamefont {Spreng}}, \bibinfo {author}
  {\bibfnamefont {N.}~\bibnamefont {Engheta}},\ and\ \bibinfo {author}
  {\bibfnamefont {J.~N.}\ \bibnamefont {Munday}},\ }\bibfield  {title}
  {\bibinfo {title} {Radiative energy band gap of nanostructures coupled with
  quantum emitters around the epsilon-near-zero frequency},\ }\href@noop {}
  {\bibfield  {journal} {\bibinfo  {journal} {Physical Review B}\ }\textbf
  {\bibinfo {volume} {106}},\ \bibinfo {pages} {085422} (\bibinfo {year}
  {2022})}\BibitemShut {NoStop}%
\bibitem [{\citenamefont {Engheta}(2013)}]{engheta2013pursuing}%
  \BibitemOpen
  \bibfield  {author} {\bibinfo {author} {\bibfnamefont {N.}~\bibnamefont
  {Engheta}},\ }\bibfield  {title} {\bibinfo {title} {Pursuing near-zero
  response},\ }\href@noop {} {\bibfield  {journal} {\bibinfo  {journal}
  {Science}\ }\textbf {\bibinfo {volume} {340}},\ \bibinfo {pages} {286}
  (\bibinfo {year} {2013})}\BibitemShut {NoStop}%
\bibitem [{\citenamefont {Ciattoni}\ \emph {et~al.}(2013)\citenamefont
  {Ciattoni}, \citenamefont {Marini}, \citenamefont {Rizza}, \citenamefont
  {Scalora},\ and\ \citenamefont {Biancalana}}]{ciattoni2013polariton}%
  \BibitemOpen
  \bibfield  {author} {\bibinfo {author} {\bibfnamefont {A.}~\bibnamefont
  {Ciattoni}}, \bibinfo {author} {\bibfnamefont {A.}~\bibnamefont {Marini}},
  \bibinfo {author} {\bibfnamefont {C.}~\bibnamefont {Rizza}}, \bibinfo
  {author} {\bibfnamefont {M.}~\bibnamefont {Scalora}},\ and\ \bibinfo {author}
  {\bibfnamefont {F.}~\bibnamefont {Biancalana}},\ }\bibfield  {title}
  {\bibinfo {title} {Polariton excitation in epsilon-near-zero slabs: Transient
  trapping of slow light},\ }\href@noop {} {\bibfield  {journal} {\bibinfo
  {journal} {Physical Review A}\ }\textbf {\bibinfo {volume} {87}},\ \bibinfo
  {pages} {053853} (\bibinfo {year} {2013})}\BibitemShut {NoStop}%
\bibitem [{\citenamefont {Javani}\ and\ \citenamefont
  {Stockman}(2016)}]{javani2016real}%
  \BibitemOpen
  \bibfield  {author} {\bibinfo {author} {\bibfnamefont {M.~H.}\ \bibnamefont
  {Javani}}\ and\ \bibinfo {author} {\bibfnamefont {M.~I.}\ \bibnamefont
  {Stockman}},\ }\bibfield  {title} {\bibinfo {title} {Real and imaginary
  properties of epsilon-near-zero materials},\ }\href@noop {} {\bibfield
  {journal} {\bibinfo  {journal} {Physical review letters}\ }\textbf {\bibinfo
  {volume} {117}},\ \bibinfo {pages} {107404} (\bibinfo {year}
  {2016})}\BibitemShut {NoStop}%
\bibitem [{\citenamefont {Raki{\'c}}(1995)}]{rakic1995algorithm}%
  \BibitemOpen
  \bibfield  {author} {\bibinfo {author} {\bibfnamefont {A.~D.}\ \bibnamefont
  {Raki{\'c}}},\ }\bibfield  {title} {\bibinfo {title} {Algorithm for the
  determination of intrinsic optical constants of metal films: application to
  aluminum},\ }\href@noop {} {\bibfield  {journal} {\bibinfo  {journal}
  {Applied optics}\ }\textbf {\bibinfo {volume} {34}},\ \bibinfo {pages} {4755}
  (\bibinfo {year} {1995})}\BibitemShut {NoStop}%
\bibitem [{\citenamefont {Silvestri}\ \emph {et~al.}(2024)\citenamefont
  {Silvestri}, \citenamefont {Sahoo}, \citenamefont {Assogna}, \citenamefont
  {Benassi}, \citenamefont {Ferrante}, \citenamefont {Ciattoni},\ and\
  \citenamefont {Marini}}]{silvestri2024resonant}%
  \BibitemOpen
  \bibfield  {author} {\bibinfo {author} {\bibfnamefont {M.}~\bibnamefont
  {Silvestri}}, \bibinfo {author} {\bibfnamefont {A.}~\bibnamefont {Sahoo}},
  \bibinfo {author} {\bibfnamefont {L.}~\bibnamefont {Assogna}}, \bibinfo
  {author} {\bibfnamefont {P.}~\bibnamefont {Benassi}}, \bibinfo {author}
  {\bibfnamefont {C.}~\bibnamefont {Ferrante}}, \bibinfo {author}
  {\bibfnamefont {A.}~\bibnamefont {Ciattoni}},\ and\ \bibinfo {author}
  {\bibfnamefont {A.}~\bibnamefont {Marini}},\ }\bibfield  {title} {\bibinfo
  {title} {Resonant third-harmonic generation driven by out-of-equilibrium
  electron dynamics in sodium-based near-zero index thin films},\ }\href@noop
  {} {\bibfield  {journal} {\bibinfo  {journal} {Nanophotonics}\ } (\bibinfo
  {year} {2024})}\BibitemShut {NoStop}%
\bibitem [{\citenamefont {Sachet}\ \emph {et~al.}(2015)\citenamefont {Sachet},
  \citenamefont {Shelton}, \citenamefont {Harris}, \citenamefont {Gaddy},
  \citenamefont {Irving}, \citenamefont {Curtarolo}, \citenamefont {Donovan},
  \citenamefont {Hopkins}, \citenamefont {Sharma}, \citenamefont {Sharma} \emph
  {et~al.}}]{sachet2015dysprosium}%
  \BibitemOpen
  \bibfield  {author} {\bibinfo {author} {\bibfnamefont {E.}~\bibnamefont
  {Sachet}}, \bibinfo {author} {\bibfnamefont {C.~T.}\ \bibnamefont {Shelton}},
  \bibinfo {author} {\bibfnamefont {J.~S.}\ \bibnamefont {Harris}}, \bibinfo
  {author} {\bibfnamefont {B.~E.}\ \bibnamefont {Gaddy}}, \bibinfo {author}
  {\bibfnamefont {D.~L.}\ \bibnamefont {Irving}}, \bibinfo {author}
  {\bibfnamefont {S.}~\bibnamefont {Curtarolo}}, \bibinfo {author}
  {\bibfnamefont {B.~F.}\ \bibnamefont {Donovan}}, \bibinfo {author}
  {\bibfnamefont {P.~E.}\ \bibnamefont {Hopkins}}, \bibinfo {author}
  {\bibfnamefont {P.~A.}\ \bibnamefont {Sharma}}, \bibinfo {author}
  {\bibfnamefont {A.~L.}\ \bibnamefont {Sharma}}, \emph {et~al.},\ }\bibfield
  {title} {\bibinfo {title} {Dysprosium-doped cadmium oxide as a gateway
  material for mid-infrared plasmonics},\ }\href@noop {} {\bibfield  {journal}
  {\bibinfo  {journal} {Nature materials}\ }\textbf {\bibinfo {volume} {14}},\
  \bibinfo {pages} {414} (\bibinfo {year} {2015})}\BibitemShut {NoStop}%
\bibitem [{\citenamefont {Yang}\ \emph {et~al.}(2017)\citenamefont {Yang},
  \citenamefont {Kelley}, \citenamefont {Sachet}, \citenamefont {Campione},
  \citenamefont {Luk}, \citenamefont {Maria}, \citenamefont {Sinclair},\ and\
  \citenamefont {Brener}}]{yang2017femtosecond}%
  \BibitemOpen
  \bibfield  {author} {\bibinfo {author} {\bibfnamefont {Y.}~\bibnamefont
  {Yang}}, \bibinfo {author} {\bibfnamefont {K.}~\bibnamefont {Kelley}},
  \bibinfo {author} {\bibfnamefont {E.}~\bibnamefont {Sachet}}, \bibinfo
  {author} {\bibfnamefont {S.}~\bibnamefont {Campione}}, \bibinfo {author}
  {\bibfnamefont {T.~S.}\ \bibnamefont {Luk}}, \bibinfo {author} {\bibfnamefont
  {J.-P.}\ \bibnamefont {Maria}}, \bibinfo {author} {\bibfnamefont {M.~B.}\
  \bibnamefont {Sinclair}},\ and\ \bibinfo {author} {\bibfnamefont
  {I.}~\bibnamefont {Brener}},\ }\bibfield  {title} {\bibinfo {title}
  {Femtosecond optical polarization switching using a cadmium oxide-based
  perfect absorber},\ }\href@noop {} {\bibfield  {journal} {\bibinfo  {journal}
  {Nature Photonics}\ }\textbf {\bibinfo {volume} {11}},\ \bibinfo {pages}
  {390} (\bibinfo {year} {2017})}\BibitemShut {NoStop}%
\bibitem [{\citenamefont {Wang}\ \emph {et~al.}(2015)\citenamefont {Wang},
  \citenamefont {Capretti},\ and\ \citenamefont {Dal~Negro}}]{wang2015wide}%
  \BibitemOpen
  \bibfield  {author} {\bibinfo {author} {\bibfnamefont {Y.}~\bibnamefont
  {Wang}}, \bibinfo {author} {\bibfnamefont {A.}~\bibnamefont {Capretti}},\
  and\ \bibinfo {author} {\bibfnamefont {L.}~\bibnamefont {Dal~Negro}},\
  }\bibfield  {title} {\bibinfo {title} {Wide tuning of the optical and
  structural properties of alternative plasmonic materials},\ }\href@noop {}
  {\bibfield  {journal} {\bibinfo  {journal} {Optical Materials Express}\
  }\textbf {\bibinfo {volume} {5}},\ \bibinfo {pages} {2415} (\bibinfo {year}
  {2015})}\BibitemShut {NoStop}%
\bibitem [{\citenamefont {Kim}\ \emph {et~al.}(2024)\citenamefont {Kim},
  \citenamefont {Kim}, \citenamefont {Jeon}, \citenamefont {Lee}, \citenamefont
  {Lee}, \citenamefont {Kim}, \citenamefont {Lee}, \citenamefont {Kim},\ and\
  \citenamefont {Kim}}]{kim2024perovskite}%
  \BibitemOpen
  \bibfield  {author} {\bibinfo {author} {\bibfnamefont {H.}~\bibnamefont
  {Kim}}, \bibinfo {author} {\bibfnamefont {G.}~\bibnamefont {Kim}}, \bibinfo
  {author} {\bibfnamefont {Y.-U.}\ \bibnamefont {Jeon}}, \bibinfo {author}
  {\bibfnamefont {W.}~\bibnamefont {Lee}}, \bibinfo {author} {\bibfnamefont
  {B.-H.}\ \bibnamefont {Lee}}, \bibinfo {author} {\bibfnamefont {I.~S.}\
  \bibnamefont {Kim}}, \bibinfo {author} {\bibfnamefont {K.}~\bibnamefont
  {Lee}}, \bibinfo {author} {\bibfnamefont {S.~J.}\ \bibnamefont {Kim}},\ and\
  \bibinfo {author} {\bibfnamefont {J.}~\bibnamefont {Kim}},\ }\bibfield
  {title} {\bibinfo {title} {Perovskite lanthanum-doped barium stannate: A
  refractory near-zero-index material for high-temperature energy harvesting
  systems},\ }\href@noop {} {\bibfield  {journal} {\bibinfo  {journal}
  {Advanced Science}\ }\textbf {\bibinfo {volume} {11}},\ \bibinfo {pages}
  {2302410} (\bibinfo {year} {2024})}\BibitemShut {NoStop}%
\bibitem [{\citenamefont {Han}\ \emph {et~al.}(2024)\citenamefont {Han},
  \citenamefont {Qiu}, \citenamefont {Liu}, \citenamefont {Chen}, \citenamefont
  {Liang}, \citenamefont {Yuan}, \citenamefont {Du},\ and\ \citenamefont
  {Ye}}]{han2024tunable}%
  \BibitemOpen
  \bibfield  {author} {\bibinfo {author} {\bibfnamefont {C.}~\bibnamefont
  {Han}}, \bibinfo {author} {\bibfnamefont {J.}~\bibnamefont {Qiu}}, \bibinfo
  {author} {\bibfnamefont {H.}~\bibnamefont {Liu}}, \bibinfo {author}
  {\bibfnamefont {K.}~\bibnamefont {Chen}}, \bibinfo {author} {\bibfnamefont
  {S.}~\bibnamefont {Liang}}, \bibinfo {author} {\bibfnamefont
  {J.}~\bibnamefont {Yuan}}, \bibinfo {author} {\bibfnamefont {M.}~\bibnamefont
  {Du}},\ and\ \bibinfo {author} {\bibfnamefont {H.}~\bibnamefont {Ye}},\
  }\bibfield  {title} {\bibinfo {title} {Tunable enz properties in organic
  material pedot: Pss treated with different solutions},\ }\href@noop {}
  {\bibfield  {journal} {\bibinfo  {journal} {Optical Materials Express}\
  }\textbf {\bibinfo {volume} {14}},\ \bibinfo {pages} {1631} (\bibinfo {year}
  {2024})}\BibitemShut {NoStop}%
\bibitem [{\citenamefont {Lee}\ \emph {et~al.}(2018)\citenamefont {Lee},
  \citenamefont {Garoni}, \citenamefont {Kita}, \citenamefont {Kamada},
  \citenamefont {Woo}, \citenamefont {Jun}, \citenamefont {Chae}, \citenamefont
  {Kim}, \citenamefont {Lee}, \citenamefont {Yoon} \emph
  {et~al.}}]{lee2018strong}%
  \BibitemOpen
  \bibfield  {author} {\bibinfo {author} {\bibfnamefont {Y.~U.}\ \bibnamefont
  {Lee}}, \bibinfo {author} {\bibfnamefont {E.}~\bibnamefont {Garoni}},
  \bibinfo {author} {\bibfnamefont {H.}~\bibnamefont {Kita}}, \bibinfo {author}
  {\bibfnamefont {K.}~\bibnamefont {Kamada}}, \bibinfo {author} {\bibfnamefont
  {B.~H.}\ \bibnamefont {Woo}}, \bibinfo {author} {\bibfnamefont {Y.~C.}\
  \bibnamefont {Jun}}, \bibinfo {author} {\bibfnamefont {S.~M.}\ \bibnamefont
  {Chae}}, \bibinfo {author} {\bibfnamefont {H.~J.}\ \bibnamefont {Kim}},
  \bibinfo {author} {\bibfnamefont {K.~J.}\ \bibnamefont {Lee}}, \bibinfo
  {author} {\bibfnamefont {S.}~\bibnamefont {Yoon}}, \emph {et~al.},\
  }\bibfield  {title} {\bibinfo {title} {Strong nonlinear optical response in
  the visible spectral range with epsilon-near-zero organic thin films},\
  }\href@noop {} {\bibfield  {journal} {\bibinfo  {journal} {Advanced Optical
  Materials}\ }\textbf {\bibinfo {volume} {6}},\ \bibinfo {pages} {1701400}
  (\bibinfo {year} {2018})}\BibitemShut {NoStop}%
\bibitem [{\citenamefont {Passler}\ \emph {et~al.}(2019)\citenamefont
  {Passler}, \citenamefont {Razdolski}, \citenamefont {Katzer}, \citenamefont
  {Storm}, \citenamefont {Caldwell}, \citenamefont {Wolf},\ and\ \citenamefont
  {Paarmann}}]{passler2019second}%
  \BibitemOpen
  \bibfield  {author} {\bibinfo {author} {\bibfnamefont {N.~C.}\ \bibnamefont
  {Passler}}, \bibinfo {author} {\bibfnamefont {I.}~\bibnamefont {Razdolski}},
  \bibinfo {author} {\bibfnamefont {D.~S.}\ \bibnamefont {Katzer}}, \bibinfo
  {author} {\bibfnamefont {D.~F.}\ \bibnamefont {Storm}}, \bibinfo {author}
  {\bibfnamefont {J.~D.}\ \bibnamefont {Caldwell}}, \bibinfo {author}
  {\bibfnamefont {M.}~\bibnamefont {Wolf}},\ and\ \bibinfo {author}
  {\bibfnamefont {A.}~\bibnamefont {Paarmann}},\ }\bibfield  {title} {\bibinfo
  {title} {Second harmonic generation from phononic epsilon-near-zero berreman
  modes in ultrathin polar crystal films},\ }\href@noop {} {\bibfield
  {journal} {\bibinfo  {journal} {Acs Photonics}\ }\textbf {\bibinfo {volume}
  {6}},\ \bibinfo {pages} {1365} (\bibinfo {year} {2019})}\BibitemShut
  {NoStop}%
\bibitem [{\citenamefont {Li}\ \emph {et~al.}(2023)\citenamefont {Li},
  \citenamefont {Liu}, \citenamefont {Wu}, \citenamefont {Li}, \citenamefont
  {Liu},\ and\ \citenamefont {Zhong}}]{li2023broadband}%
  \BibitemOpen
  \bibfield  {author} {\bibinfo {author} {\bibfnamefont {J.}~\bibnamefont
  {Li}}, \bibinfo {author} {\bibfnamefont {S.}~\bibnamefont {Liu}}, \bibinfo
  {author} {\bibfnamefont {S.}~\bibnamefont {Wu}}, \bibinfo {author}
  {\bibfnamefont {W.}~\bibnamefont {Li}}, \bibinfo {author} {\bibfnamefont
  {Y.}~\bibnamefont {Liu}},\ and\ \bibinfo {author} {\bibfnamefont
  {Z.}~\bibnamefont {Zhong}},\ }\bibfield  {title} {\bibinfo {title} {Broadband
  absorption based on multi-layered enz film: from directional to
  omnidirectional absorption},\ }in\ \href@noop {} {\emph {\bibinfo {booktitle}
  {Journal of Physics: Conference Series}}},\ Vol.\ \bibinfo {volume} {2548}\
  (\bibinfo {organization} {IOP Publishing},\ \bibinfo {year} {2023})\ p.\
  \bibinfo {pages} {012016}\BibitemShut {NoStop}%
\bibitem [{\citenamefont {Xu}\ \emph {et~al.}(2024)\citenamefont {Xu},
  \citenamefont {Crassee}, \citenamefont {Bechtel}, \citenamefont {Zhou},
  \citenamefont {Bercher}, \citenamefont {Korosec}, \citenamefont {Rischau},
  \citenamefont {Teyssier}, \citenamefont {Crust}, \citenamefont {Lee} \emph
  {et~al.}}]{xu2024highly}%
  \BibitemOpen
  \bibfield  {author} {\bibinfo {author} {\bibfnamefont {R.}~\bibnamefont
  {Xu}}, \bibinfo {author} {\bibfnamefont {I.}~\bibnamefont {Crassee}},
  \bibinfo {author} {\bibfnamefont {H.~A.}\ \bibnamefont {Bechtel}}, \bibinfo
  {author} {\bibfnamefont {Y.}~\bibnamefont {Zhou}}, \bibinfo {author}
  {\bibfnamefont {A.}~\bibnamefont {Bercher}}, \bibinfo {author} {\bibfnamefont
  {L.}~\bibnamefont {Korosec}}, \bibinfo {author} {\bibfnamefont {C.~W.}\
  \bibnamefont {Rischau}}, \bibinfo {author} {\bibfnamefont {J.}~\bibnamefont
  {Teyssier}}, \bibinfo {author} {\bibfnamefont {K.~J.}\ \bibnamefont {Crust}},
  \bibinfo {author} {\bibfnamefont {Y.}~\bibnamefont {Lee}}, \emph {et~al.},\
  }\bibfield  {title} {\bibinfo {title} {Highly confined epsilon-near-zero and
  surface phonon polaritons in srtio3 membranes},\ }\href@noop {} {\bibfield
  {journal} {\bibinfo  {journal} {Nature Communications}\ }\textbf {\bibinfo
  {volume} {15}},\ \bibinfo {pages} {4743} (\bibinfo {year}
  {2024})}\BibitemShut {NoStop}%
\bibitem [{\citenamefont {Caldwell}\ \emph {et~al.}(2014)\citenamefont
  {Caldwell}, \citenamefont {Kretinin}, \citenamefont {Chen}, \citenamefont
  {Giannini}, \citenamefont {Fogler}, \citenamefont {Francescato},
  \citenamefont {Ellis}, \citenamefont {Tischler}, \citenamefont {Woods},
  \citenamefont {Giles} \emph {et~al.}}]{caldwell2014sub}%
  \BibitemOpen
  \bibfield  {author} {\bibinfo {author} {\bibfnamefont {J.~D.}\ \bibnamefont
  {Caldwell}}, \bibinfo {author} {\bibfnamefont {A.~V.}\ \bibnamefont
  {Kretinin}}, \bibinfo {author} {\bibfnamefont {Y.}~\bibnamefont {Chen}},
  \bibinfo {author} {\bibfnamefont {V.}~\bibnamefont {Giannini}}, \bibinfo
  {author} {\bibfnamefont {M.~M.}\ \bibnamefont {Fogler}}, \bibinfo {author}
  {\bibfnamefont {Y.}~\bibnamefont {Francescato}}, \bibinfo {author}
  {\bibfnamefont {C.~T.}\ \bibnamefont {Ellis}}, \bibinfo {author}
  {\bibfnamefont {J.~G.}\ \bibnamefont {Tischler}}, \bibinfo {author}
  {\bibfnamefont {C.~R.}\ \bibnamefont {Woods}}, \bibinfo {author}
  {\bibfnamefont {A.~J.}\ \bibnamefont {Giles}}, \emph {et~al.},\ }\bibfield
  {title} {\bibinfo {title} {Sub-diffractional volume-confined polaritons in
  the natural hyperbolic material hexagonal boron nitride},\ }\href@noop {}
  {\bibfield  {journal} {\bibinfo  {journal} {Nature communications}\ }\textbf
  {\bibinfo {volume} {5}},\ \bibinfo {pages} {5221} (\bibinfo {year}
  {2014})}\BibitemShut {NoStop}%
\bibitem [{\citenamefont {Jacob}(2014)}]{jacob2014hyperbolic}%
  \BibitemOpen
  \bibfield  {author} {\bibinfo {author} {\bibfnamefont {Z.}~\bibnamefont
  {Jacob}},\ }\bibfield  {title} {\bibinfo {title} {Hyperbolic
  phonon--polaritons},\ }\href@noop {} {\bibfield  {journal} {\bibinfo
  {journal} {Nature materials}\ }\textbf {\bibinfo {volume} {13}},\ \bibinfo
  {pages} {1081} (\bibinfo {year} {2014})}\BibitemShut {NoStop}%
\bibitem [{SM()}]{SM}%
  \BibitemOpen
  \bibfield  {title} {\bibinfo {title} {See supplemental material at [url will
  be inserted by publisher] for details on finite element modelling and
  extended discussions on the effects of damping factor},\ }\href@noop {} {\
  }\BibitemShut {NoStop}%
\bibitem [{\citenamefont {Shi}\ and\ \citenamefont {Lu}(2016)}]{shi2016field}%
  \BibitemOpen
  \bibfield  {author} {\bibinfo {author} {\bibfnamefont {K.}~\bibnamefont
  {Shi}}\ and\ \bibinfo {author} {\bibfnamefont {Z.}~\bibnamefont {Lu}},\
  }\bibfield  {title} {\bibinfo {title} {Field-effect optical modulation based
  on epsilon-near-zero conductive oxide},\ }\href@noop {} {\bibfield  {journal}
  {\bibinfo  {journal} {Optics Communications}\ }\textbf {\bibinfo {volume}
  {370}},\ \bibinfo {pages} {22} (\bibinfo {year} {2016})}\BibitemShut
  {NoStop}%
\bibitem [{\citenamefont {Mambra}\ \emph {et~al.}(2021)\citenamefont {Mambra},
  \citenamefont {Pant},\ and\ \citenamefont {Mitra}}]{mambra2021dynamic}%
  \BibitemOpen
  \bibfield  {author} {\bibinfo {author} {\bibfnamefont {A.}~\bibnamefont
  {Mambra}}, \bibinfo {author} {\bibfnamefont {R.}~\bibnamefont {Pant}},\ and\
  \bibinfo {author} {\bibfnamefont {J.}~\bibnamefont {Mitra}},\ }\bibfield
  {title} {\bibinfo {title} {Dynamic tuning of enz region of ito and sensing
  using a tapered optical fiber},\ }in\ \href@noop {} {\emph {\bibinfo
  {booktitle} {International Workshop on the Physics of Semiconductor and
  Devices}}}\ (\bibinfo {organization} {Springer},\ \bibinfo {year} {2021})\
  pp.\ \bibinfo {pages} {229--236}\BibitemShut {NoStop}%
\bibitem [{\citenamefont {Feigenbaum}\ \emph {et~al.}(2010)\citenamefont
  {Feigenbaum}, \citenamefont {Diest},\ and\ \citenamefont
  {Atwater}}]{feigenbaum2010unity}%
  \BibitemOpen
  \bibfield  {author} {\bibinfo {author} {\bibfnamefont {E.}~\bibnamefont
  {Feigenbaum}}, \bibinfo {author} {\bibfnamefont {K.}~\bibnamefont {Diest}},\
  and\ \bibinfo {author} {\bibfnamefont {H.~A.}\ \bibnamefont {Atwater}},\
  }\bibfield  {title} {\bibinfo {title} {Unity-order index change in
  transparent conducting oxides at visible frequencies},\ }\href@noop {}
  {\bibfield  {journal} {\bibinfo  {journal} {Nano letters}\ }\textbf {\bibinfo
  {volume} {10}},\ \bibinfo {pages} {2111} (\bibinfo {year}
  {2010})}\BibitemShut {NoStop}%
\bibitem [{\citenamefont {Loudon}(1970)}]{loudon1970propagation}%
  \BibitemOpen
  \bibfield  {author} {\bibinfo {author} {\bibfnamefont {R.}~\bibnamefont
  {Loudon}},\ }\bibfield  {title} {\bibinfo {title} {The propagation of
  electromagnetic energy through an absorbing dielectric},\ }\href@noop {}
  {\bibfield  {journal} {\bibinfo  {journal} {Journal of Physics A: General
  Physics}\ }\textbf {\bibinfo {volume} {3}},\ \bibinfo {pages} {233} (\bibinfo
  {year} {1970})}\BibitemShut {NoStop}%
\bibitem [{\citenamefont {Brillouin}(2013)}]{brillouin2013wave}%
  \BibitemOpen
  \bibfield  {author} {\bibinfo {author} {\bibfnamefont {L.}~\bibnamefont
  {Brillouin}},\ }\href@noop {} {\emph {\bibinfo {title} {Wave propagation and
  group velocity}}},\ Vol.~\bibinfo {volume} {8}\ (\bibinfo  {publisher}
  {Academic press},\ \bibinfo {year} {2013})\BibitemShut {NoStop}%
\bibitem [{\citenamefont {Nunes}\ \emph {et~al.}(2011)\citenamefont {Nunes},
  \citenamefont {Vasconcelos}, \citenamefont {Bezerra},\ and\ \citenamefont
  {Weiner}}]{nunes2011electromagnetic}%
  \BibitemOpen
  \bibfield  {author} {\bibinfo {author} {\bibfnamefont {F.~D.}\ \bibnamefont
  {Nunes}}, \bibinfo {author} {\bibfnamefont {T.~C.}\ \bibnamefont
  {Vasconcelos}}, \bibinfo {author} {\bibfnamefont {M.}~\bibnamefont
  {Bezerra}},\ and\ \bibinfo {author} {\bibfnamefont {J.}~\bibnamefont
  {Weiner}},\ }\bibfield  {title} {\bibinfo {title} {Electromagnetic energy
  density in dispersive and dissipative media},\ }\href@noop {} {\bibfield
  {journal} {\bibinfo  {journal} {Journal of the Optical Society of America B}\
  }\textbf {\bibinfo {volume} {28}},\ \bibinfo {pages} {1544} (\bibinfo {year}
  {2011})}\BibitemShut {NoStop}%
\bibitem [{\citenamefont {Reshef}\ \emph {et~al.}(2019)\citenamefont {Reshef},
  \citenamefont {De~Leon}, \citenamefont {Alam},\ and\ \citenamefont
  {Boyd}}]{reshef2019nonlinear}%
  \BibitemOpen
  \bibfield  {author} {\bibinfo {author} {\bibfnamefont {O.}~\bibnamefont
  {Reshef}}, \bibinfo {author} {\bibfnamefont {I.}~\bibnamefont {De~Leon}},
  \bibinfo {author} {\bibfnamefont {M.~Z.}\ \bibnamefont {Alam}},\ and\
  \bibinfo {author} {\bibfnamefont {R.~W.}\ \bibnamefont {Boyd}},\ }\bibfield
  {title} {\bibinfo {title} {Nonlinear optical effects in epsilon-near-zero
  media},\ }\href@noop {} {\bibfield  {journal} {\bibinfo  {journal} {Nature
  Reviews Materials}\ }\textbf {\bibinfo {volume} {4}},\ \bibinfo {pages} {535}
  (\bibinfo {year} {2019})}\BibitemShut {NoStop}%
\bibitem [{\citenamefont {Huang}\ \emph {et~al.}(2016)\citenamefont {Huang},
  \citenamefont {Lee}, \citenamefont {Sokhoyan}, \citenamefont {Pala},
  \citenamefont {Thyagarajan}, \citenamefont {Han}, \citenamefont {Tsai},\ and\
  \citenamefont {Atwater}}]{huang2016gate}%
  \BibitemOpen
  \bibfield  {author} {\bibinfo {author} {\bibfnamefont {Y.-W.}\ \bibnamefont
  {Huang}}, \bibinfo {author} {\bibfnamefont {H.~W.~H.}\ \bibnamefont {Lee}},
  \bibinfo {author} {\bibfnamefont {R.}~\bibnamefont {Sokhoyan}}, \bibinfo
  {author} {\bibfnamefont {R.~A.}\ \bibnamefont {Pala}}, \bibinfo {author}
  {\bibfnamefont {K.}~\bibnamefont {Thyagarajan}}, \bibinfo {author}
  {\bibfnamefont {S.}~\bibnamefont {Han}}, \bibinfo {author} {\bibfnamefont
  {D.~P.}\ \bibnamefont {Tsai}},\ and\ \bibinfo {author} {\bibfnamefont
  {H.~A.}\ \bibnamefont {Atwater}},\ }\bibfield  {title} {\bibinfo {title}
  {Gate-tunable conducting oxide metasurfaces},\ }\href@noop {} {\bibfield
  {journal} {\bibinfo  {journal} {Nano letters}\ }\textbf {\bibinfo {volume}
  {16}},\ \bibinfo {pages} {5319} (\bibinfo {year} {2016})}\BibitemShut
  {NoStop}%
\bibitem [{\citenamefont {Maniyara}\ \emph {et~al.}(2019)\citenamefont
  {Maniyara}, \citenamefont {Rodrigo}, \citenamefont {Yu}, \citenamefont
  {Canet-Ferrer}, \citenamefont {Ghosh}, \citenamefont {Yongsunthon},
  \citenamefont {Baker}, \citenamefont {Rezikyan}, \citenamefont
  {Garc{\'\i}a~de Abajo},\ and\ \citenamefont {Pruneri}}]{maniyara2019tunable}%
  \BibitemOpen
  \bibfield  {author} {\bibinfo {author} {\bibfnamefont {R.~A.}\ \bibnamefont
  {Maniyara}}, \bibinfo {author} {\bibfnamefont {D.}~\bibnamefont {Rodrigo}},
  \bibinfo {author} {\bibfnamefont {R.}~\bibnamefont {Yu}}, \bibinfo {author}
  {\bibfnamefont {J.}~\bibnamefont {Canet-Ferrer}}, \bibinfo {author}
  {\bibfnamefont {D.~S.}\ \bibnamefont {Ghosh}}, \bibinfo {author}
  {\bibfnamefont {R.}~\bibnamefont {Yongsunthon}}, \bibinfo {author}
  {\bibfnamefont {D.~E.}\ \bibnamefont {Baker}}, \bibinfo {author}
  {\bibfnamefont {A.}~\bibnamefont {Rezikyan}}, \bibinfo {author}
  {\bibfnamefont {F.~J.}\ \bibnamefont {Garc{\'\i}a~de Abajo}},\ and\ \bibinfo
  {author} {\bibfnamefont {V.}~\bibnamefont {Pruneri}},\ }\bibfield  {title}
  {\bibinfo {title} {Tunable plasmons in ultrathin metal films},\ }\href@noop
  {} {\bibfield  {journal} {\bibinfo  {journal} {Nature Photonics}\ }\textbf
  {\bibinfo {volume} {13}},\ \bibinfo {pages} {328} (\bibinfo {year}
  {2019})}\BibitemShut {NoStop}%
\bibitem [{\citenamefont {Milonni}(2004)}]{milonni2004fast}%
  \BibitemOpen
  \bibfield  {author} {\bibinfo {author} {\bibfnamefont {P.~W.}\ \bibnamefont
  {Milonni}},\ }\href@noop {} {\emph {\bibinfo {title} {Fast light, slow light
  and left-handed light}}}\ (\bibinfo  {publisher} {CRC Press},\ \bibinfo
  {year} {2004})\BibitemShut {NoStop}%
\bibitem [{\citenamefont {Jackson}\ and\ \citenamefont
  {Fox}(1999)}]{jackson1999classical}%
  \BibitemOpen
  \bibfield  {author} {\bibinfo {author} {\bibfnamefont {J.~D.}\ \bibnamefont
  {Jackson}}\ and\ \bibinfo {author} {\bibfnamefont {R.~F.}\ \bibnamefont
  {Fox}},\ }\href@noop {} {\bibinfo {title} {Classical electrodynamics}}
  (\bibinfo {year} {1999})\BibitemShut {NoStop}%
\end{thebibliography}%

\onecolumngrid
\appendix
\section{Supplementary Material}
\subsection{Dielectric permittivity of ITO, CdO and PEDOT:PSS}
\begin{figure}[h]
	\begin{center}
		\includegraphics[width=10cm]{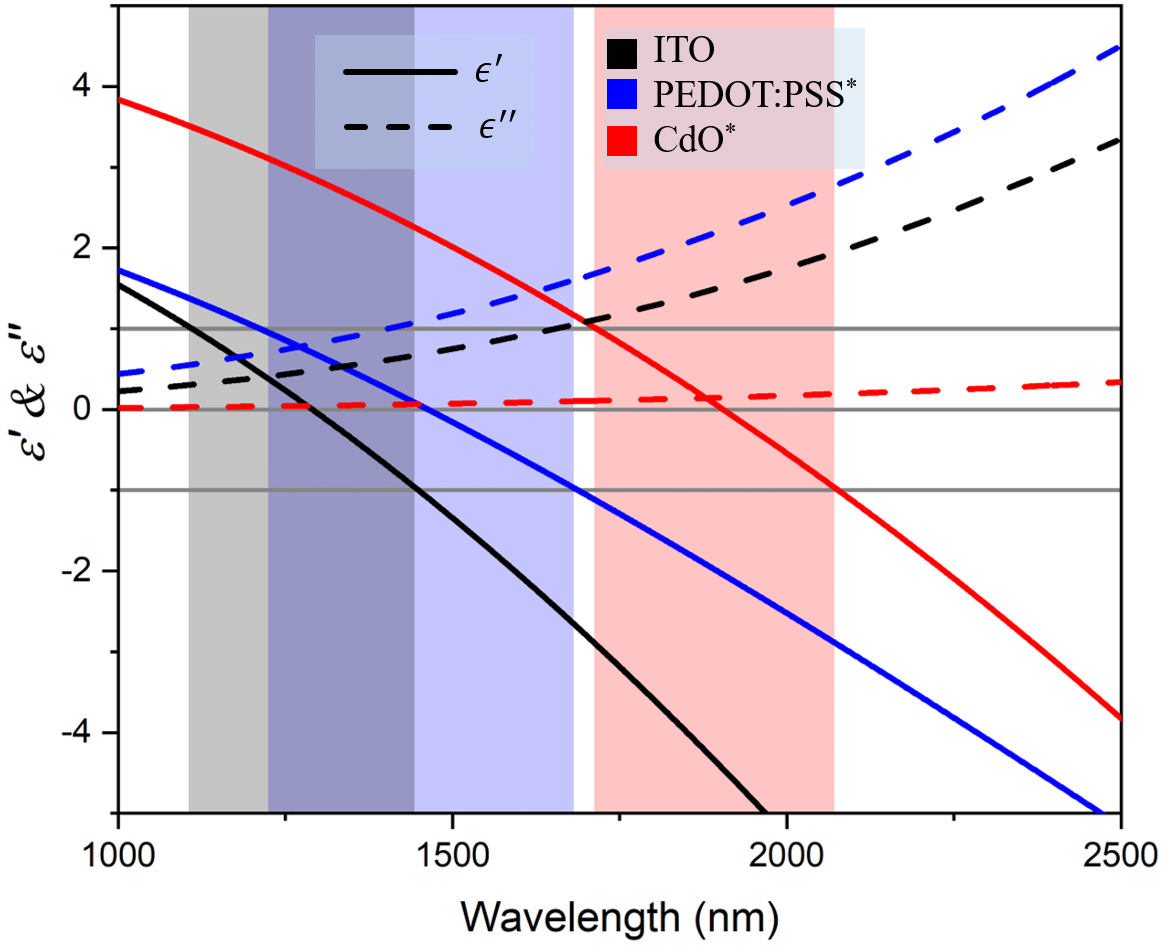}
		\caption{Real ($\epsilon'$) and imaginary ($\epsilon''$) part of dielectric permittivity for ITO, CdO and PEDOT:PSS plotted as a function of wavelength. The colored regions demarcate the ENZ regime i.e. the spectral regime within which $|\epsilon'|<1$, for each material. * - doped}
		\label{fig:eps}
	\end{center}
\end{figure}

\subsection{Finite element Method Modeling}

All finite element simulations were carried out using the Wave Optics and Semiconductor modules of COMSOL Multiphysics (5.3a). The wave-optics simulations were performed with Perfectly Matched Layer (PML) boundary conditions applied around the simulation geometry. PML boundary condition is most preferred here since it allows to simulate open boundaries by fully absorbing outgoing wave, thus removing reflections from the boundary. A PML layer thickness (1 $\mu$m) of the order of excitation wavelength was used for maximum absorption at the boundary. The mesh density for ultra-thin (5 nm) ENZ films was optimized with free triangular pattern to have a mesh size ranging from 1 $\AA$ to 0.1 $\AA$. On the other hand, while using semiconductor module to gate the ENZ layer, mapped mesh pattern with a geometric sequence distribution was used, since high mesh density ($<$0.1 $\AA$) around the gate interface was essential to resolve the change in free carrier density and refractive index across depth of the ENZ layer. Other geometric domains were meshed with simple free triangular pattern and PML were meshed with mapped pattern.

\subsubsection{Wave-optics module}
\begin{figure}[h]
	\begin{center}
		\includegraphics[width=12cm]{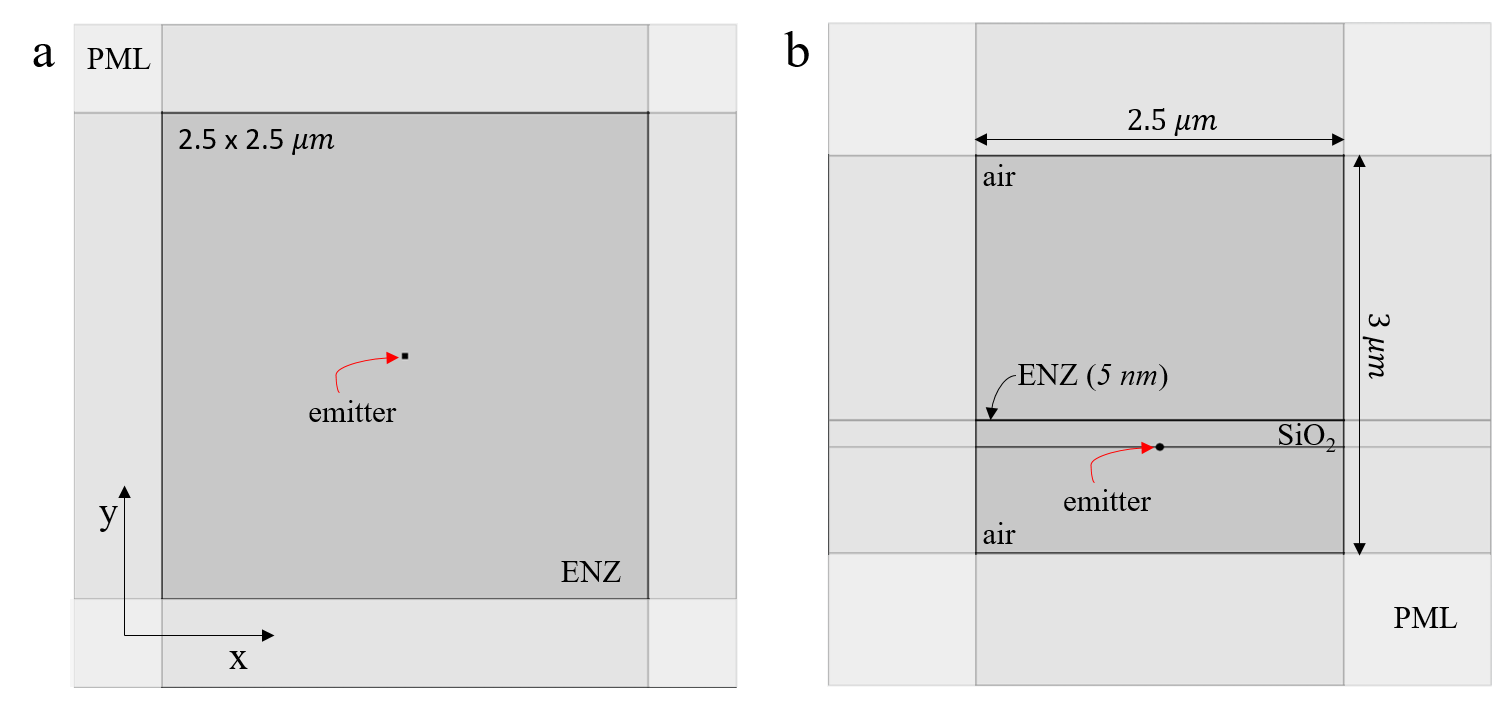}
		\caption{Simulation geometry for (a) emitter embedded in Unbounded ENZ media, and (b) dynamic control of radiation transmission using ultra-thin film of ENZ.}
		\label{fig:geo}
	\end{center}
\end{figure}

Fig. \ref{fig:geo} shows the two simulation geometry with the emitter embedded in unbounded ENZ media (fig. \ref{fig:geo}a) and dynamic control of radiation transmission through ENZ media (fig. \ref{fig:geo}b).
The ENZ media under wave-optics module was defined using Drude model and the parameters used are as shown in table \ref{ENZwave}. The point emitter was simulated using electric point dipole node in wave optics module with a dipole moment: $p = \omega \cdot 1\times$ 10$^{-10} ~ [A\cdot m]$,  oriented in $x$ axis. The emitter is encapsulated in a vacuum bubble (dia: 50 nm) to save the solutions from diverging due to presence of source and sink at the same point. In the beam steering simulations a dipole emitter array was mimicked using surface current density node with $J = \omega \cdot 5 \times 10^{-5}$ [A/m].

\begin{table}[h!]
	\centering
	\begin{tabular}{|c|c|}
		\hline
		\textbf{Property} & \textbf{Value} \\ \hline
		High Frequency Permittivity & 3.9 \\ \hline
		Free carrier Density (N$_c$) & 6 $\times$ 10$^{26}$ m$^{-3}$ \\ \hline
		Effect mass of carrier & 0.35 m$_e$ \\ \hline
		Scattering rate ($\gamma$) & $10^{10}$ - $10^{14}$ Hz \\ \hline
	\end{tabular}
	\caption{Material parameters defining ENZ media in Wave optics module. $m_e$ is the free electron mass}
	\label{ENZwave}
\end{table}

\subsubsection{Semiconductor Module}

Thin insulated gate node of the semiconductor module and $HfO_2$ layer with $~1$ nm thickness was modelled to gate the ENZ media and Fermi-Dirac carrier statistic was used to solve for number density of majority carriers. The material parameters used to describe the ENZ and $HfO_2$ layer in semiconductor module is tabulated in table \ref{ENZsemi}. In semiconductor and wave-optics coupled simulations, the semiconductor module is first used to solve the variation in number density (N$_c$) in the ENZ layer as a function of gate voltage (V$_G$). This simulated number density across the bulk of ENZ layer is fed into the Drude model which effectively calculates the refractive index ($\tilde{n}(x,y)=n+i\kappa$) across the bulk of ENZ media. This $\tilde{n}(x,y)$ is further used by wave-optics model to solve wave equations.

\begin{table}[h!]
	\centering
	\begin{tabular}{|c|c|c|}
		\hline
		\textbf{Property} & \textbf{ENZ} & \textbf{HfO$_2$} \\ \hline
		Relative permittivity & 9.3 & 20 \\ \hline
		Band gap & 3.5 eV & 11 eV \\ \hline
		Electron affinity & 4.1 eV & 0.1 eV \\ \hline
		Effective Density of states (VB) & 1.0236 $\times 10^{25}$ m$^{-3}$ & - \\ \hline
		Effective Density of states (CB) & 5.196 $\times 10^{24}$ m$^{-3}$ & - \\ \hline
		Electron Mobility & 23 cm$^2$ V$^{-1}$ s$^{-1}$ & - \\ \hline
		Hole Mobility & 1 cm$^2$ V$^{-1}$ s$^{-1}$ &  - \\ \hline
		%Row 8, Col 1 & Row 8, Col 2 \\ \hline
	\end{tabular}
	\caption{Materials properties used in the semiconductor module}
	\label{ENZsemi}
\end{table}

\subsection{Non-radiative modes and electromagnetic wave impedance in ENZ media}	
\begin{figure}[h]
	\begin{center}
		\includegraphics[width=12cm]{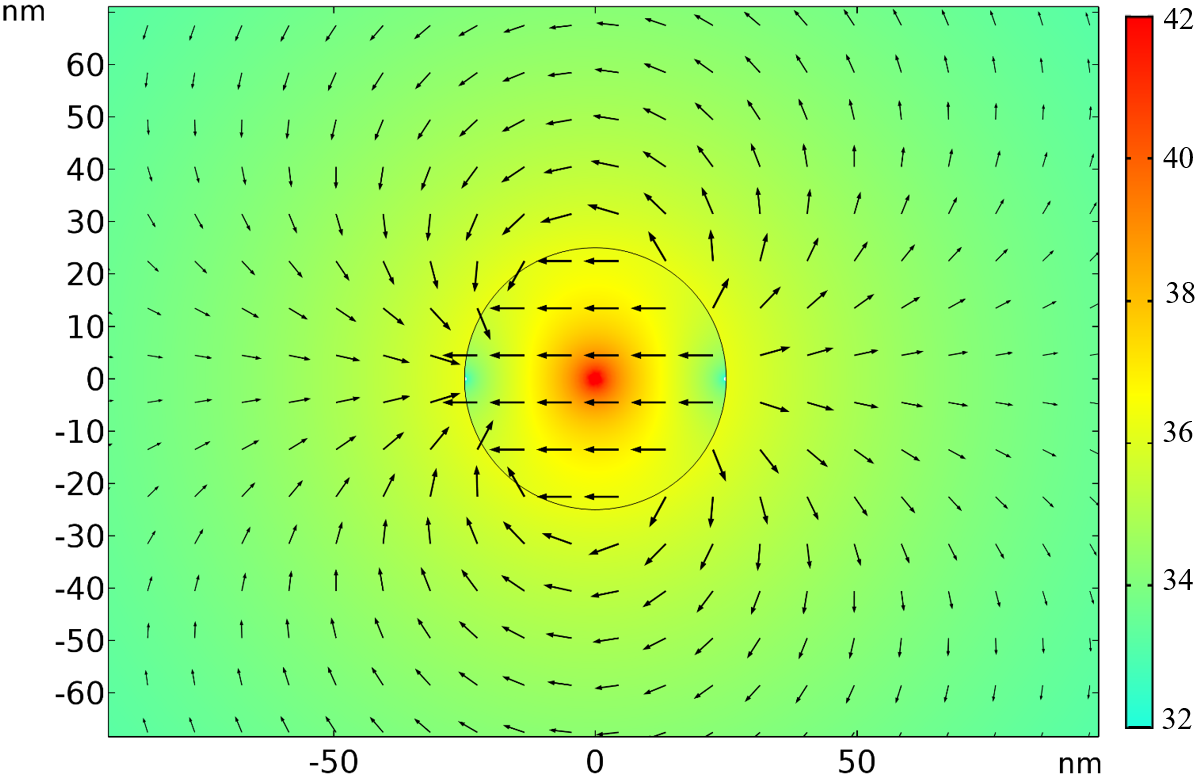}
		\caption{Electrostatic dipolar field for $\lambda_{ENZ}$=1600 nm and $\lambda_{em}$=1600 nm}
		\label{fig:electrostatic}
	\end{center}
\end{figure}

When, $\lambda_{em} = \lambda_{ENZ}$ and $\gamma = 10^{10}$ Hz, the dipole oscillator embedded inside the ENZ media mimics the $\vec{E}$ profile of a electrostatic dipolar field, as shown in fig. \ref{fig:electrostatic}. The $\vec{E}$ still remains oscillating inside the vacuum bubble, it only shows the electrostatic nature in the ENZ media. The excitation of such a field profile is understood as non-radiative mode. The false color in the background corresponds to $log(|\vec{E}|)$ as depicted by the colorbar.

%	\subsection{Impedance of ENZ media}
\begin{figure}
	\begin{center}
		\includegraphics[width=10cm]{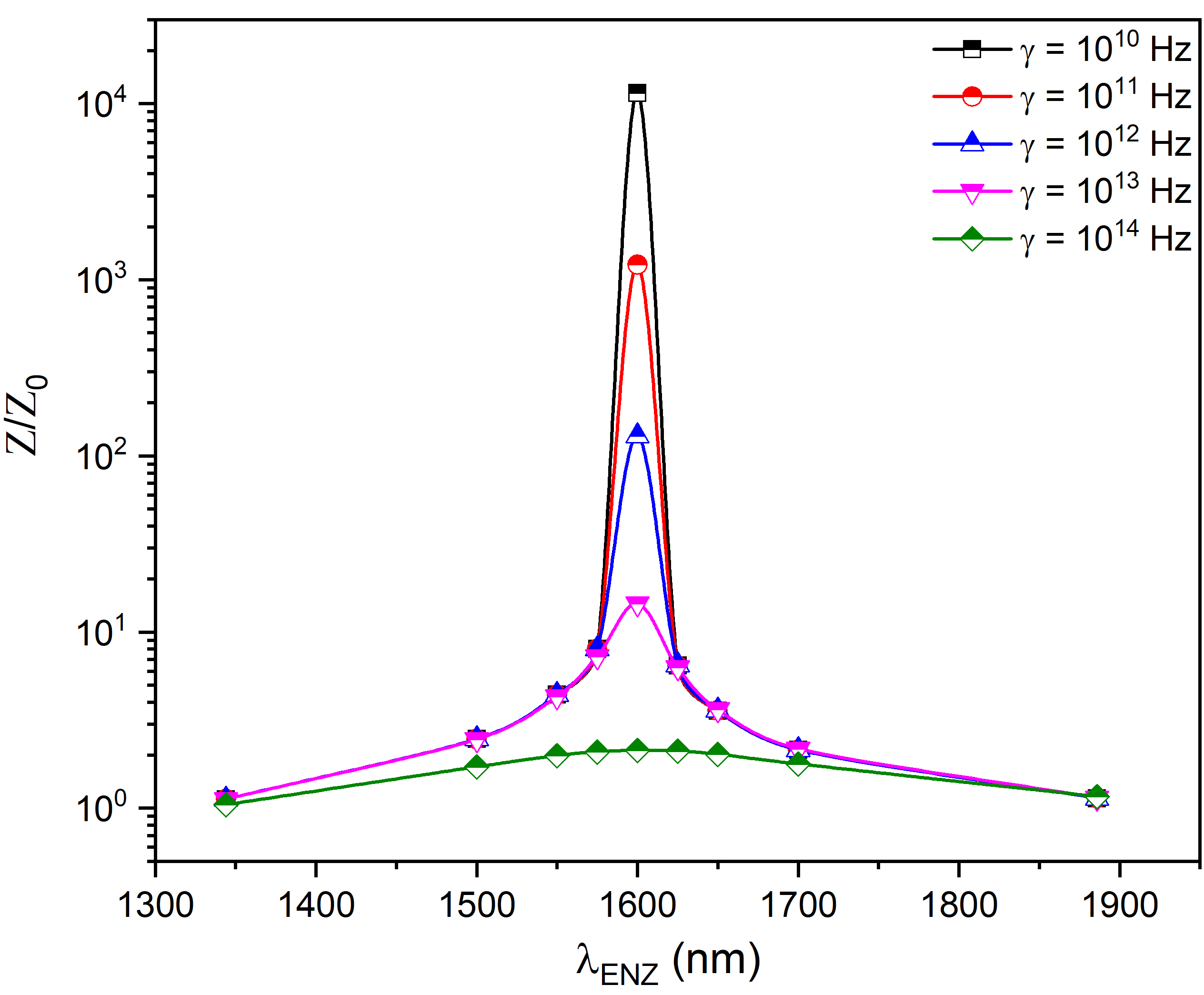}
		\caption{Normalized impedance $Z/Z_0$ vs $\lambda_{ENZ}$ for different $\gamma$.}
		\label{fig:imp}
	\end{center}
\end{figure}

Fig. \ref{fig:imp} plots the normalized electromagnetic impedance of ENZ media defined as $Z/Z_0$, where $Z = |\vec{E}|/|\vec{H}|$ and $Z_0 =$ 377 $\Omega$ is the  impedance of free space. For the embedded dipole radiating at 1600 nm, fig. \ref{fig:imp} shows that $Z$ is maximum when $\lambda_{ENZ} = \lambda_{em}$. Radiation encounters lower impedance  when the surrounding ENZ media is a dielectric or a metal. Importantly, reducing the $\gamma$, which decreases the loss in the media, leads to a pronounced increase in $Z$ as $\epsilon'\rightarrow0$. The enhanced impedance corroborates the observed suppression in transmittance when $\lambda_{ENZ}=\lambda_{em}$ (fig. 3a \& fig. 4a, main text) and further discussed below.

\subsection{Energy velocity $v_E$ in the  ENZ media}

\begin{figure}
	\begin{center}
		\includegraphics[width=0.9\linewidth]{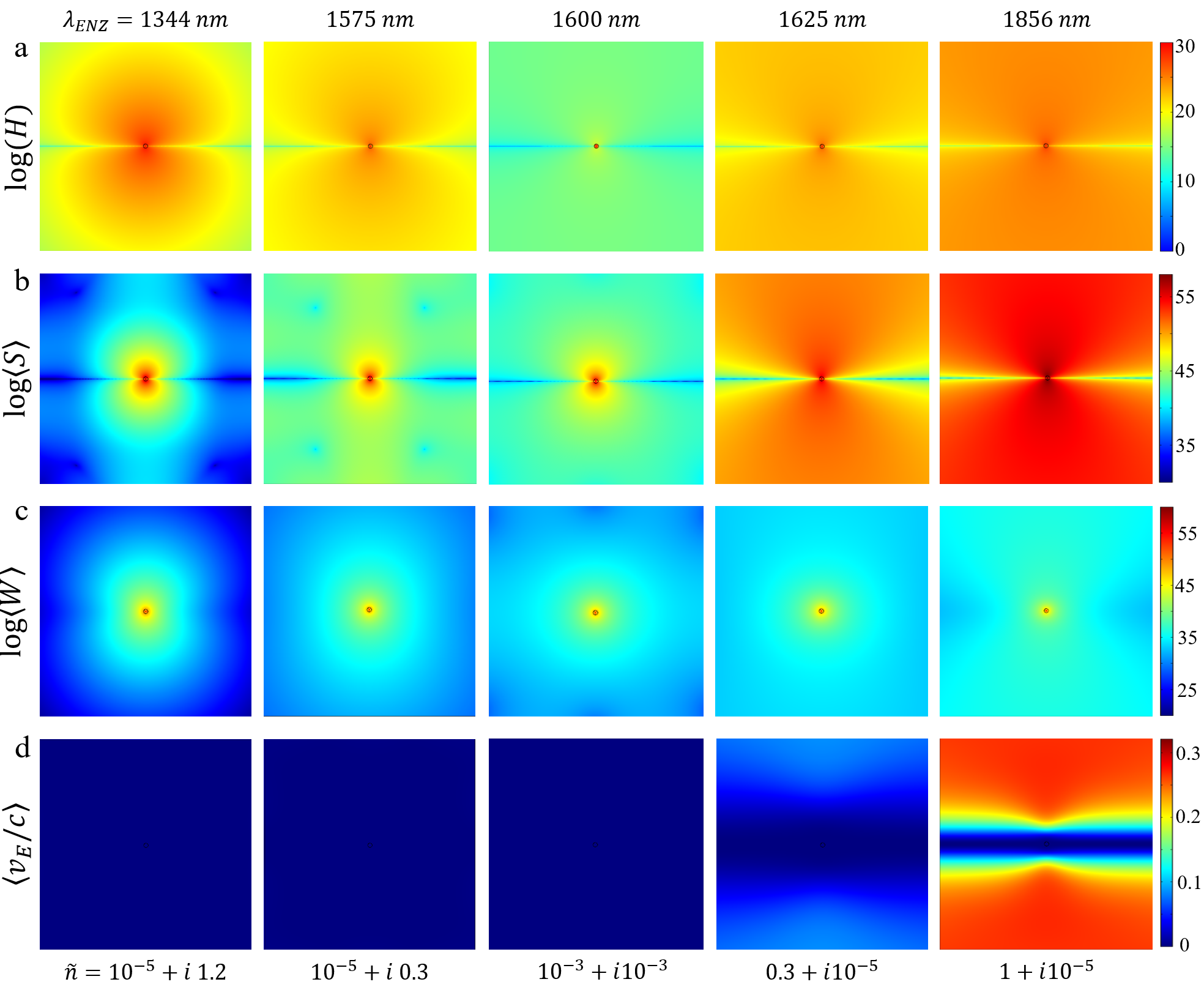}
		\caption{ Spatial variation of time-averaged (a) Magnetic field $|\vec{H}|$, (b) Poynting vector $\langle S \rangle$, (c)  energy density $\langle W \rangle$ plots in log scale inside an ENZ media with different $\lambda_{ENZ}$, with an oscillating dipole radiating at $\lambda_{em} = $ 1600 nm embedded at the centre (d) Plots the normalized, time-averaged energy velocity $v_{E}/c$. The  $\lambda_{ENZ}$ of the ENZ media and its refractive index at 1600 nm are noted on top of each column.}
		\label{fig:vE}
	\end{center}
\end{figure}

The rate at which energy and information is transmitted along with a propagating EM wave is fairly represented by the group velocity ($v_g=\partial\omega/\partial k$), if both the dispersion and loss in the medium are small, i.e. $\partial\epsilon'/\partial\omega<<\epsilon'/\omega$ and $\epsilon''<<\epsilon'$. 
Though the dielectric constant of free-carrier type ENZ media investigated here, modelled by the Drude model lack spectrally localized absorption features, both the above conditions are violated, especially in the metallic regime i.e. $\omega/\omega_z<1$. The metallic regime  also exhibits anamalous dispersion ($dn/d\omega<0$) and yields a $v_g$ that is orders of magnitude higher than $c$ (fig \ref{fig:n-dn}b), which is unphysical and no longer represents the speed at which energy or information is transmitted. In the presence of  absorption and dispersion, including anamalous dispersion, the energy velocity $v_E$, defined as the ratio of the time averaged Poynting vector to energy density of the wave i.e. $v_E  = \langle S \rangle/\langle W \rangle$ represents the rate at which energy travels in a medium \cite{brillouin2013wave,loudon1970propagation}. 
Here, $\vec{S}$ and $W$ are defined as;
\begin{equation}
	\vec{S} = \frac{1}{2} Re\{\vec{E}\times \vec{H}^*\}
\end{equation} 
\begin{equation}
	W = \frac{\epsilon_0}{4} \left(\epsilon'+\frac{2\omega\epsilon''}{\gamma}\right) |E|^2 + \frac{\mu_0}{4} |H|^2
\end{equation}
\noindent
Note that for an ideal loss-less material, $\gamma \rightarrow 0$ and $v_E \rightarrow v_g$. Fig. \ref{fig:vE} plots the spatial variation (a) magnetic field ($\vec{H}$), (b) Poynting vector $\vec{S}$, (c) energy density $W$ and (d) $v_E/c$ for a dipole radiating in ENZ media with various $\lambda_{ENZ}$ in the range 1344 nm to 1856 nm, corresponding to the electric field plots shown in fig. 2. 
The plots for $log|\vec{H}|$ (fig. \ref{fig:vE}a) show that the magnetic field in the ENZ medium with $\lambda_{ENZ} = \lambda_{em}$ drastically decreases in magnitude  due to the decoupling of the electric and magnetic field components as $\epsilon'\rightarrow0$. In the metallic regime i.e. $\lambda_{ENZ}$ = 1575 nm and 1344 nm the $\vec{H}$ recovers  but shows strong spatial decay due to the lossy nature of the medium. A similar behavior is observed in the case of the $\vec{S}$, where the radiated power quenches when $\lambda_{ENZ} = \lambda_{em}$ as shown in fig. \ref{fig:vE}b, which plots the time averaged  $log \langle |\vec{S}| \rangle$. The suppression in $\vec{S}$ in the ENZ regime is commensurate with the maximum in electromagnetic impedance discussed earlier.
The corresponding energy density plots in fig. \ref{fig:vE}c ($log \langle W \rangle$) then allows estimation of  $v_E$ in each case and is shown in fig. \ref{fig:vE}d. The normalized $\langle v_E/c \rangle$ is highest when the ENZ medium is a dielectric ($\lambda_{ENZ}$=1856 nm) at 1600 nm with a magnitude $<$ 1, which arises due to the finite loss in the system. $\langle v_E/c \rangle$ decreases monotonically as the ENZ medium transits from the dielectric to the metallic regime, across the ENZ regime. Evidently, energy propagation becomes strongly suppressed as $\lambda_{ENZ}$ approaches  $\lambda_{em}$, with further damping observed in the metallic regime, $\lambda_{ENZ} < \lambda_{em}$.

An analytical expression for $v_E$ is given as \cite{loudon1970propagation}:
\begin{equation}
	v_E = \frac{c}{n} \left(1+\frac{2\omega\kappa}{n\gamma}\right)^{-1}
\end{equation}
Fig. 3d plots the spectral variation of $v_E$ with different $\gamma$. It can be seen that as $\omega \rightarrow \infty$, $v_E \rightarrow v_p$.

\subsection{Why not Group Velocity!}
\begin{figure}[h]
	\begin{center}
		\includegraphics[width=10cm]{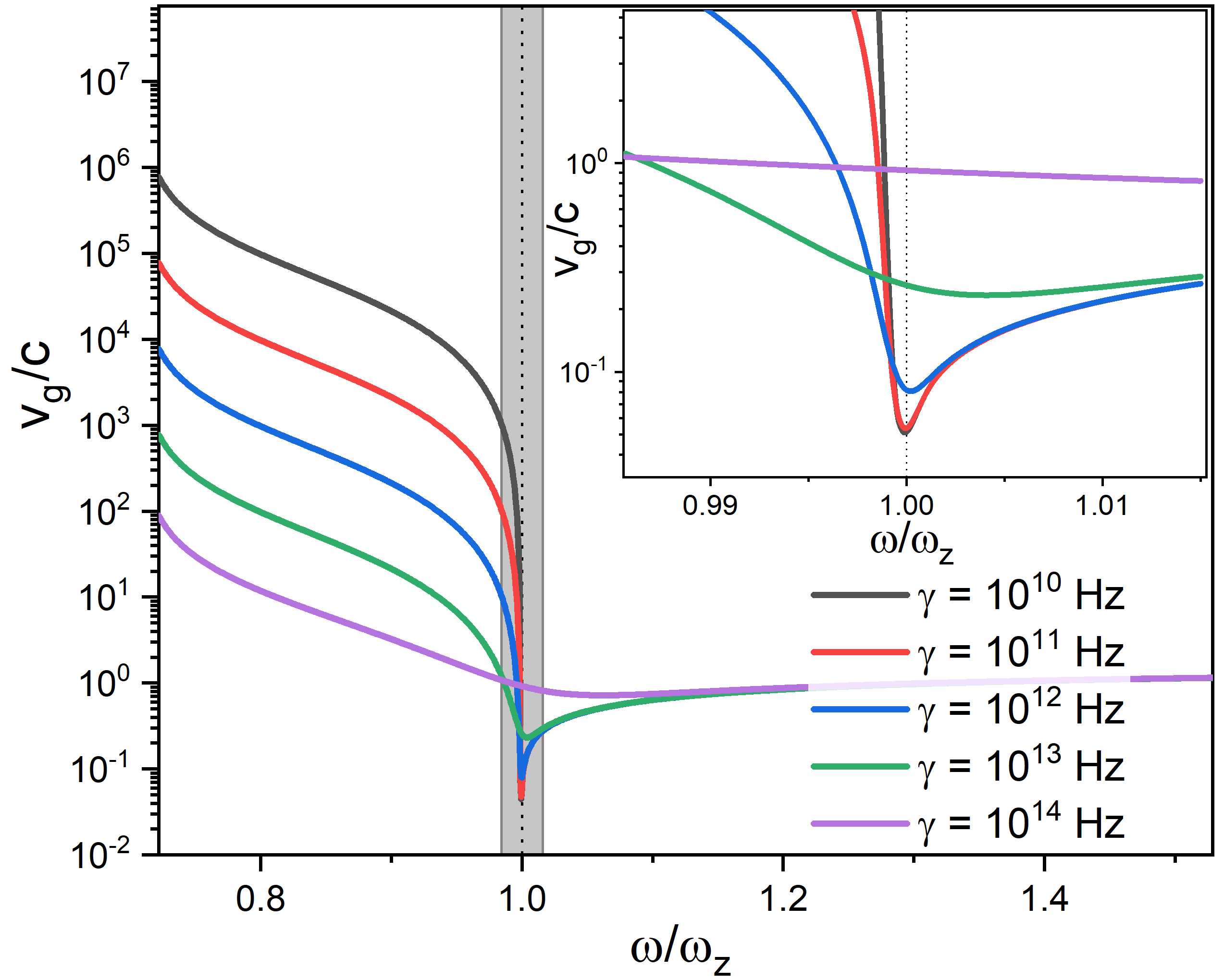}
		\caption{Spectral variation of $v_g/c$ for different $\gamma$. Inset: variation near $\omega_z$}
		\label{fig:vg}
	\end{center}
\end{figure}

The group velocity ($v_g = \frac{\partial \omega}{\partial k}$) plotted here is considering only the real part of the wavevector $\vec{k}$. For $\omega/\omega_z >> 1$ (dielectric behavior) the metric $v_g$ saturated to $v_p$, and for $\omega/\omega_z \approx 1$ the $v_g$ hits a minima  highlighting the slow-light characteristics of ENZ media \cite{liberal2017near,ciattoni2013polariton,reshef2019nonlinear}. However, for $\omega/\omega_z < 1$ the media exhibits metallic behavior with increased loss. High absorption in this regime results in superluminal values for $v_g$ \cite{milonni2004fast,jackson1999classical} and $v_g$ becomes less reliable as a metric for describing energy or information transport. In contrast, the alternative metric $v_E$ (energy velocity), as discussed above, more accurately captures the energy and information propagation across all three regimes.

\subsection{Impact of Free Carrier Density Modulation in ENZ media}
\begin{figure}[h]
	\begin{center}
		\includegraphics[width=17cm]{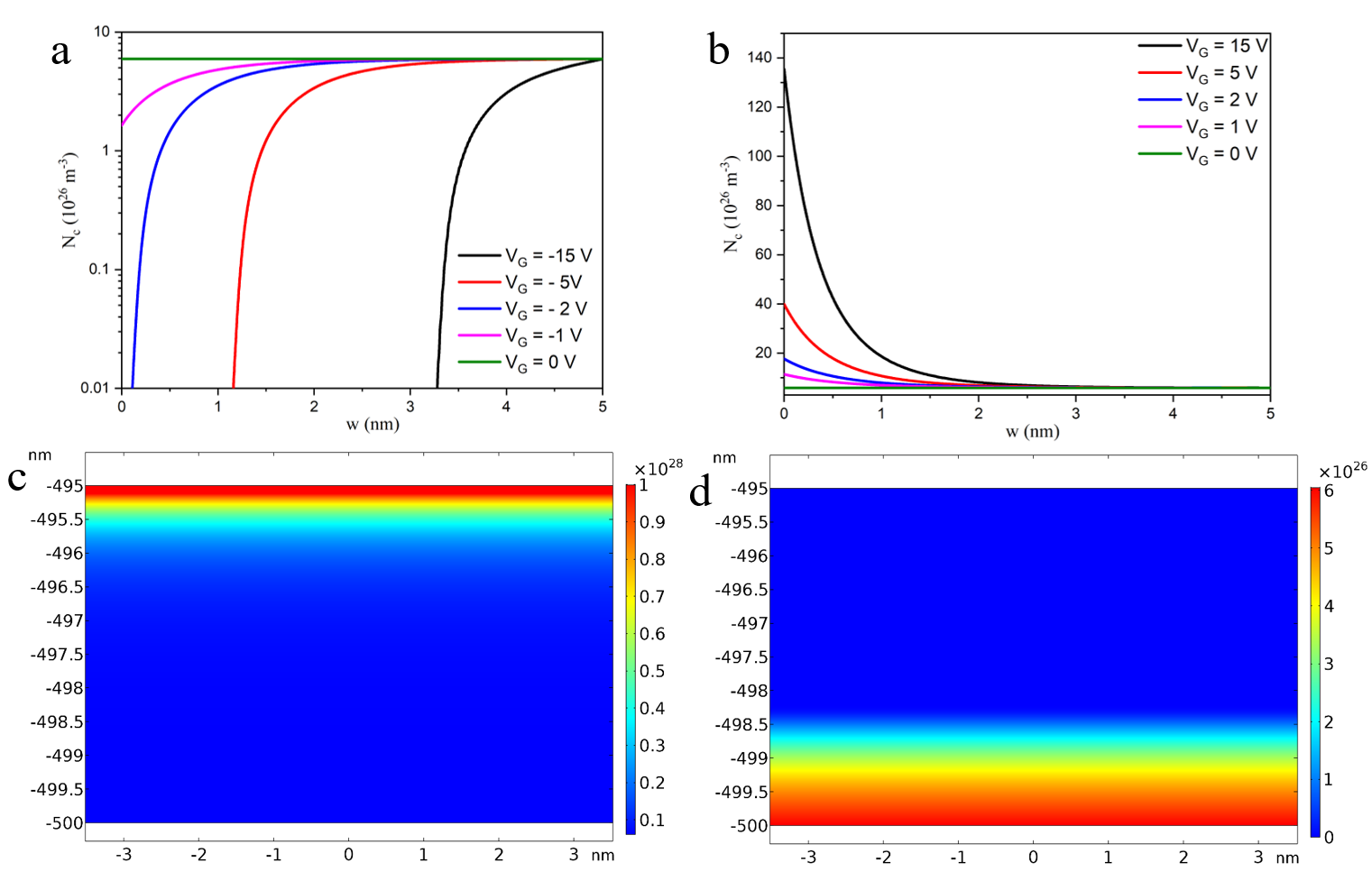}
		\caption{Variation in number density (N$_c$) for (a) Negative and (b) Positive gate voltage. 2D plot shows depletion and accumulation width for (c) V$_G$ = -15 V and (d) V$_G$ = 15 V across the ENZ media.}
		\label{fig:Nd}
	\end{center}
\end{figure}

\begin{figure}[h]
	\begin{center}
		\includegraphics[height=22cm]{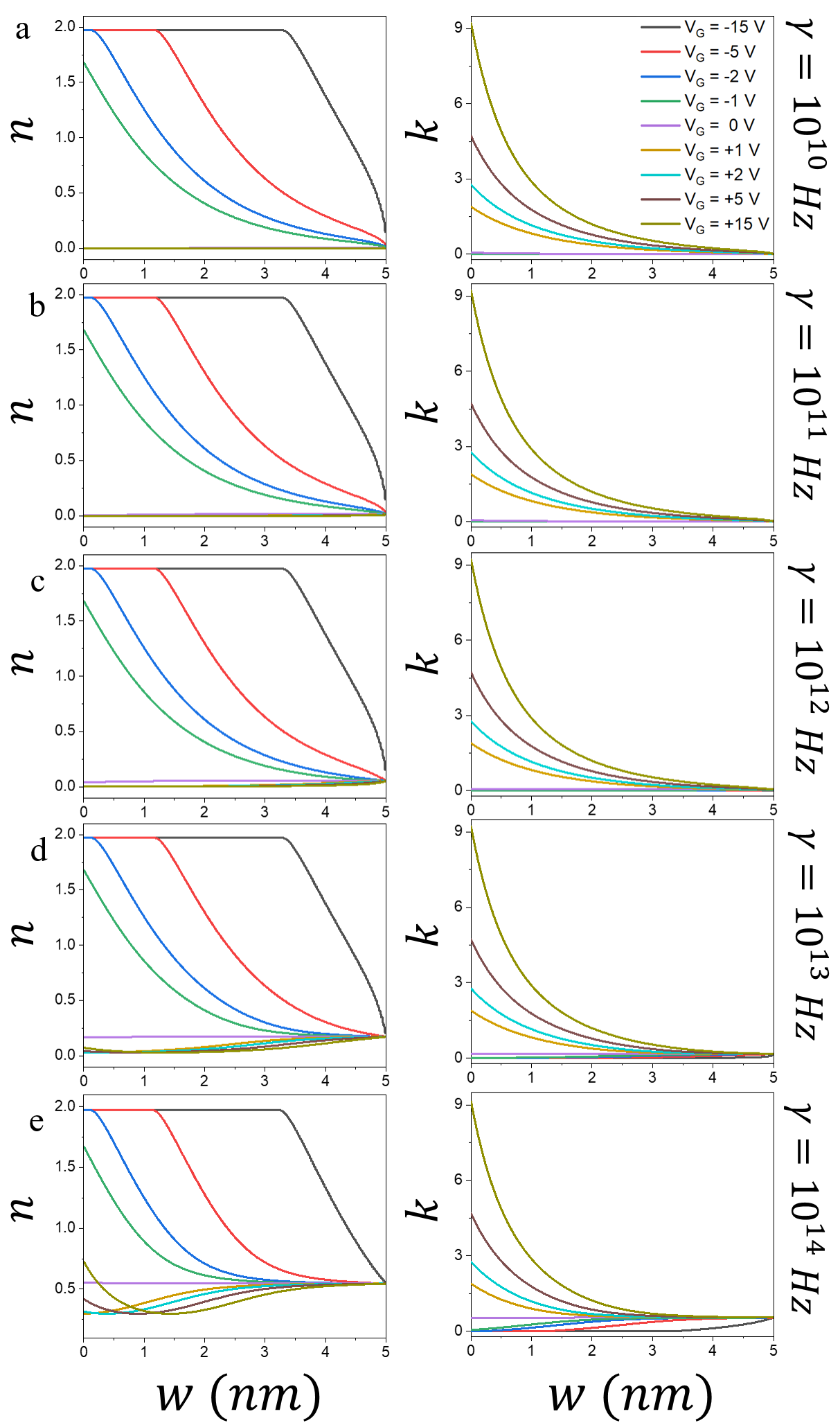}
		\caption{Calculated $n$ and $\kappa$ across the depth of the ENZ layer for varied V$_G$ and $\gamma$.}
		\label{fig:n}
	\end{center}
\end{figure}

\begin{figure}
	\begin{center}
		\includegraphics[width=14cm]{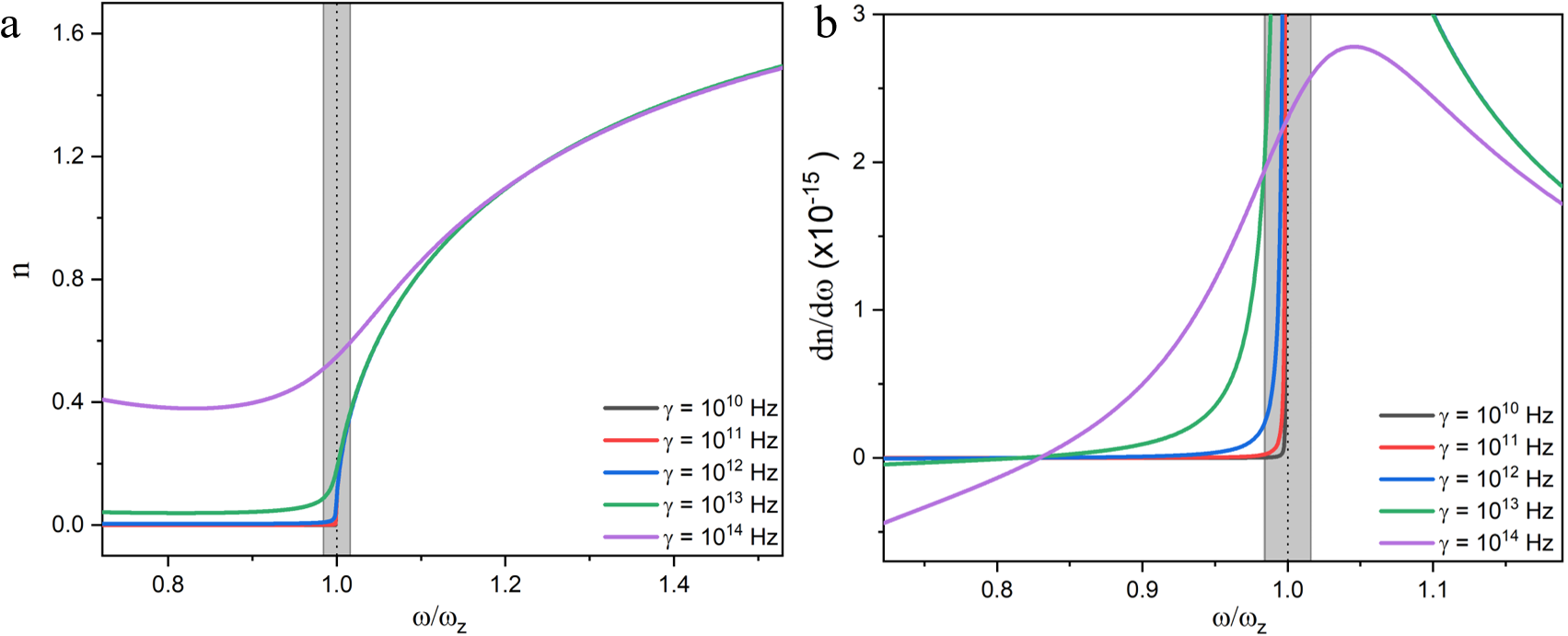}
		\caption{(a) Real refractive index ($n$) and (b) $dn/d\omega$ as a function of $\omega$ for different $\gamma$.}
		\label{fig:n-dn}
	\end{center}
\end{figure}

Fig. \ref{fig:Nd}a and \ref{fig:Nd}b shows the variation in the free-carrier number density for negative and positive (V$_G$), respectively. Similarly the 2D plots in fig. \ref{fig:Nd}c and \ref{fig:Nd}d shows the variation in N$_c$ across the whole bulk of ENZ media at -15 V and +15 V, respectively. These plots signify that V$_G$ = -15 V creates a depletion width of $\sim$ 3 nm and V$_G$ = +15 V creates an accumulation width of $\sim$ 1 nm. The real and imaginary part of refractive index (fig. \ref{fig:n}) of the ENZ media is derived from the simulated N$_c$ (fig. \ref{fig:Nd}) with different material damping ($\gamma = 10^{10} - 10^{14}$ Hz).

\subsection{Effect of $\gamma$ and film thickness}
\vspace{-1 cm}
\begin{figure}[h]
	\begin{center}
		\includegraphics[width=15cm]{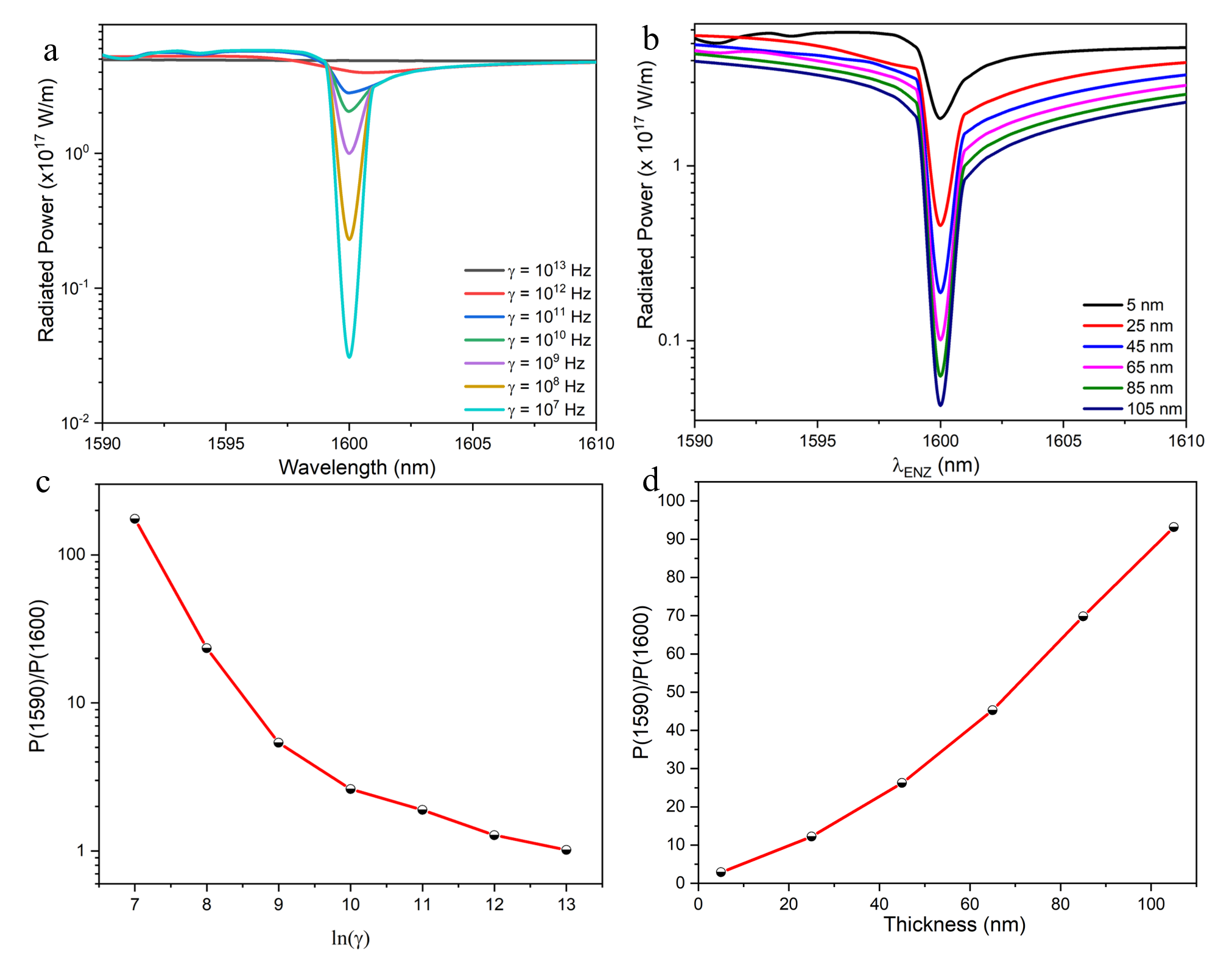}
		\caption{Effect of radiated power as a function of (a)$\gamma$, for 5 nm thick film and (b) film thickness, for $\gamma = 10^{10}$ Hz. The ON-OFF ratio is calculated with P(1590 nm)/P(1600 nm) for different (c) $\gamma$ and (d) thickness of film.}
		\label{fig:gamma_thickness}
	\end{center}
\end{figure}

\subsection{Effect of HfO$_2$ thickness}

\begin{figure}[h]
	\begin{center}
		\includegraphics[width=8cm]{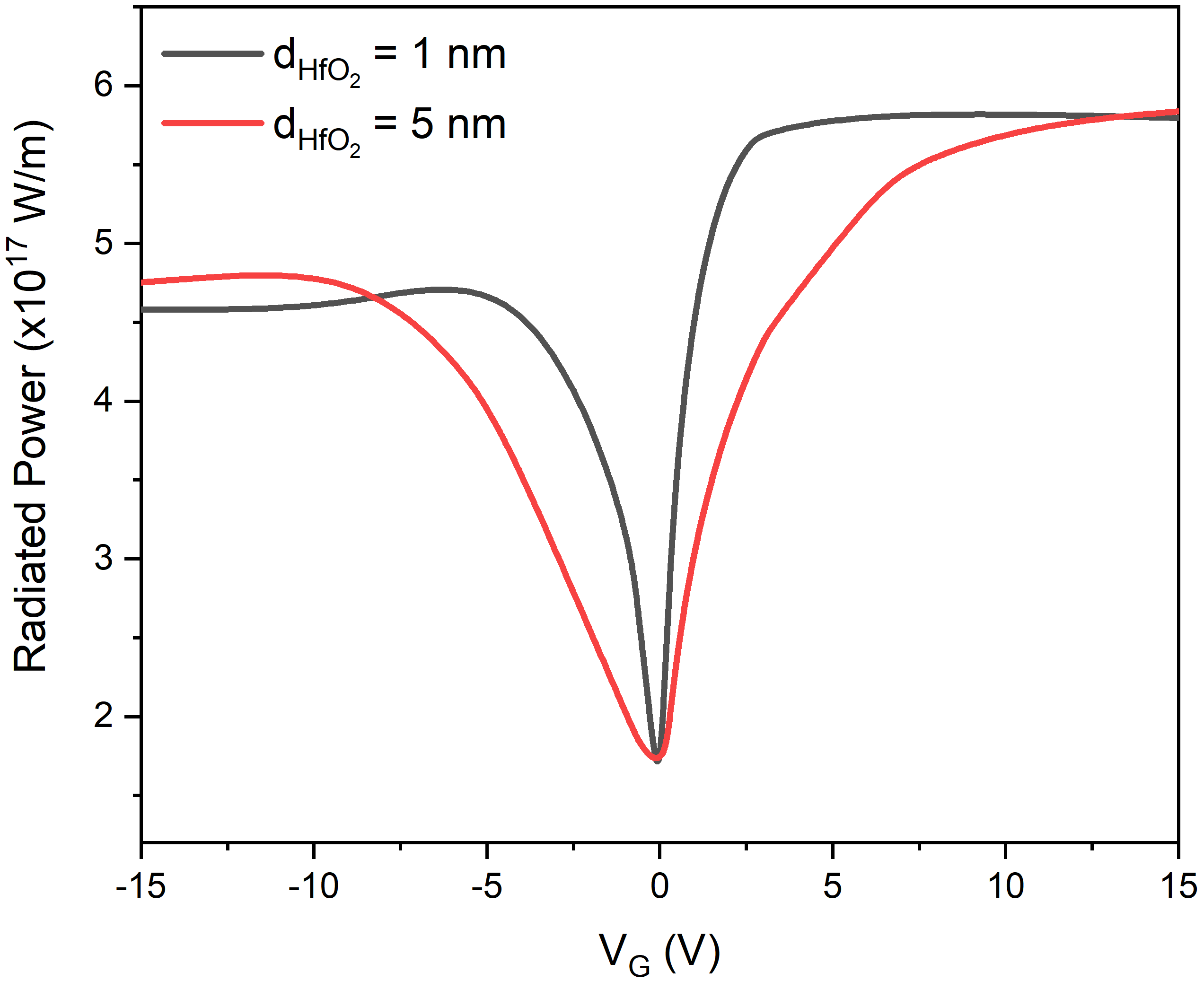}
		\caption{Comparison of radiated power transmitted through the 5nm ENZ film on SiO2 substrate with 1nm and 5nm thick HfO2 layer for different V$_G$} 
		\label{fig:hfo2}
	\end{center}
\end{figure}

In the simulations, the 1 nm HfO$_2$ thickness was used to maximize carrier depletion/accumulation. The calculations also show that thicker HfO$_2$ films can achieve comparable results by increasing the applied gate voltage. Fig. \ref{fig:hfo2} illustrates the transmission modulation as a function of V$_G$, with 5 nm HfO$_2$ gate insulator on a 5 nm ENZ thin film (fig. 4a inset), in addition to the data shown in fig. 5a of the manuscript for 1 nm thick insulator. The width of the dip in transmitted power broadens for a thicker gate insulator due to the reduced gate capacitance (see table \ref{hfo2}).

\begin{table}[h]
	\centering
	\begin{tabular}{|c|c|c|c|c|}
		\hline
		\textbf{Thickness (nm)} & \textbf{V$_G$ (V)} & \textbf{C ($\mu F/cm^2$)} & \textbf{E (V/cm)} & \textbf{D (V/cm$^2$)}\\ \hline
		1 & 2 & 17.7 & 2 x 10$^7$ & 3.54 x 10$^-3$ \\ \hline
		5 & 10 & 3.54 & 2 x 10$^7$ & 3.54 x 10$^-3$\\ \hline
	\end{tabular}
	\caption{Comparison for 1 nm and 5 nm HfO$_2$ as gate insulator. C - Capacitance, E - Electric Field, D - Induced Charge Density.}
	\label{hfo2}
\end{table}

\subsection{Effect of $\gamma$ on power flow in ENZ media}

\begin{figure}
	\begin{center}
		\includegraphics[width=15cm]{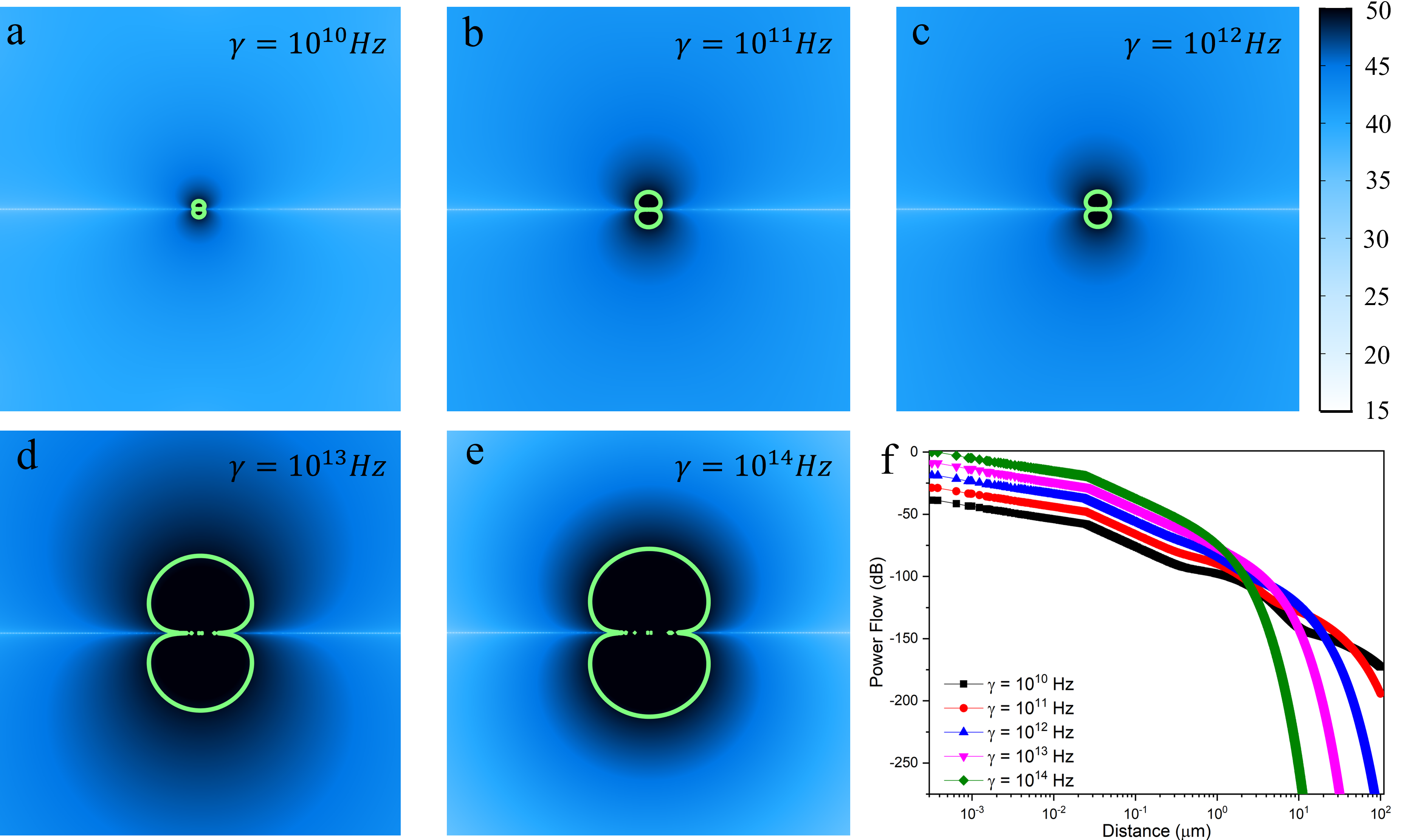}
		\caption{$|\vec{S}|$ in log scale from a point dipole embedded in unbounded ENZ media with (a) to (e) increasing $\gamma$. Green contour shows an ISO power line at 3 $\times$ 10$^{21}$ W/m$^2$. (f) Magnitude of Power in dB versus distance from the point dipole for different $\gamma$.}
		\label{fig:powerflow}
	\end{center}
\end{figure}

\begin{figure}
	\begin{center}
		\includegraphics[width=15cm]{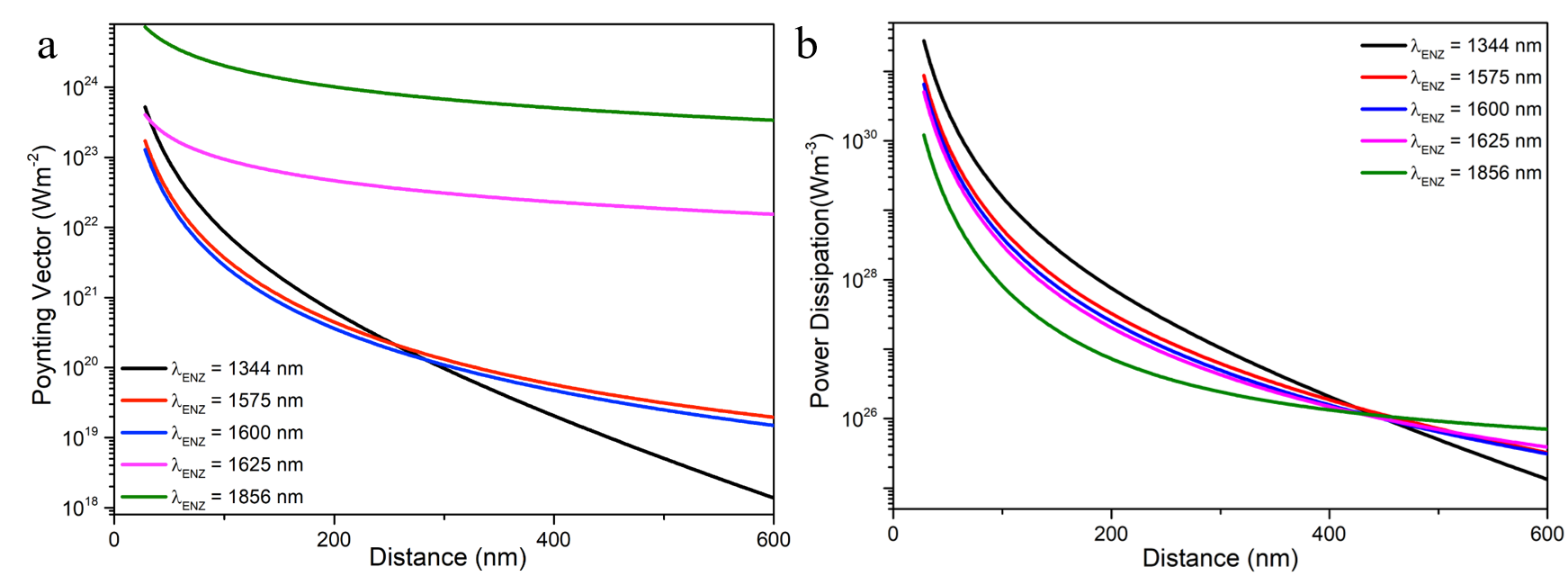}
		\caption{(a) Power flow, $|\vec{S}|$ and (b) power dissipation density as a function of distance from the point source for different $\lambda_{ENZ}$ and $\gamma = 10^{10}$ Hz.}
		\label{power_dis}
	\end{center}
\end{figure}

Power flow around a dipole embedded in an unbounded ENZ media with the $\lambda_{em}$ matching  $\lambda_{ENZ}$, exploring the role of $\gamma$ is presented in figs. \ref{fig:powerflow}a-e.
The colour map depicts a log scale plot of the magnitude of the Poynting vector ($|\vec{S}|$), with the green lines denoting iso-power flow contour of magnitude 3 $\times$ 10$^{21}$ W/m$^2$, for $\gamma$ increased from  $10^{10} - 10^{14}$ Hz.
Fig. \ref{fig:powerflow}f plots $|\vec{S}|$ (in dB) as a function of distance from the dipole emitter. 
The plots show that for the dipole emitting at the $\lambda_{ENZ}$, power coupled into the ENZ media increases with increasing ENZ damping.
Though the lowest power is coupled into the ENZ media for the smallest damping, radiation propagates less-dissipatively through the media compared to the case of higher $\gamma$. Hence, the impact of non-radiative modes at $\lambda_{ENZ}$ are more pronounced for lower $\gamma$.
Alternatively, fig. \ref{power_dis}a plots
$|\vec{S}|$	for a fixed $\gamma$ ($=10^{10}$ Hz) and variable $\lambda_{ENZ}$. Again, the power coupling into the ENZ media is least when $\lambda_{ENZ}=\lambda_{em}$ and propagates less-dissipatively. For $\lambda_{ENZ} >\lambda_{em}$ the material behaves like a dielectric, resulting in higher power coupling with least dissipation, and for $\lambda_{ENZ}<\lambda_{em}$ the media behaves metallic, which again couples in relative higher power as compared to the case of $\lambda_{ENZ}=\lambda_{em}$ but the propagation is significantly lossy. In the same scheme, fig. \ref{power_dis}b plots power dissipation density (P$_D \approx \omega \epsilon'' |\vec{E}|^2$).  

\newpage

\subsection{Multi-layer architecture}

\begin{figure}[h]
	\begin{center}
		\includegraphics[width=15cm]{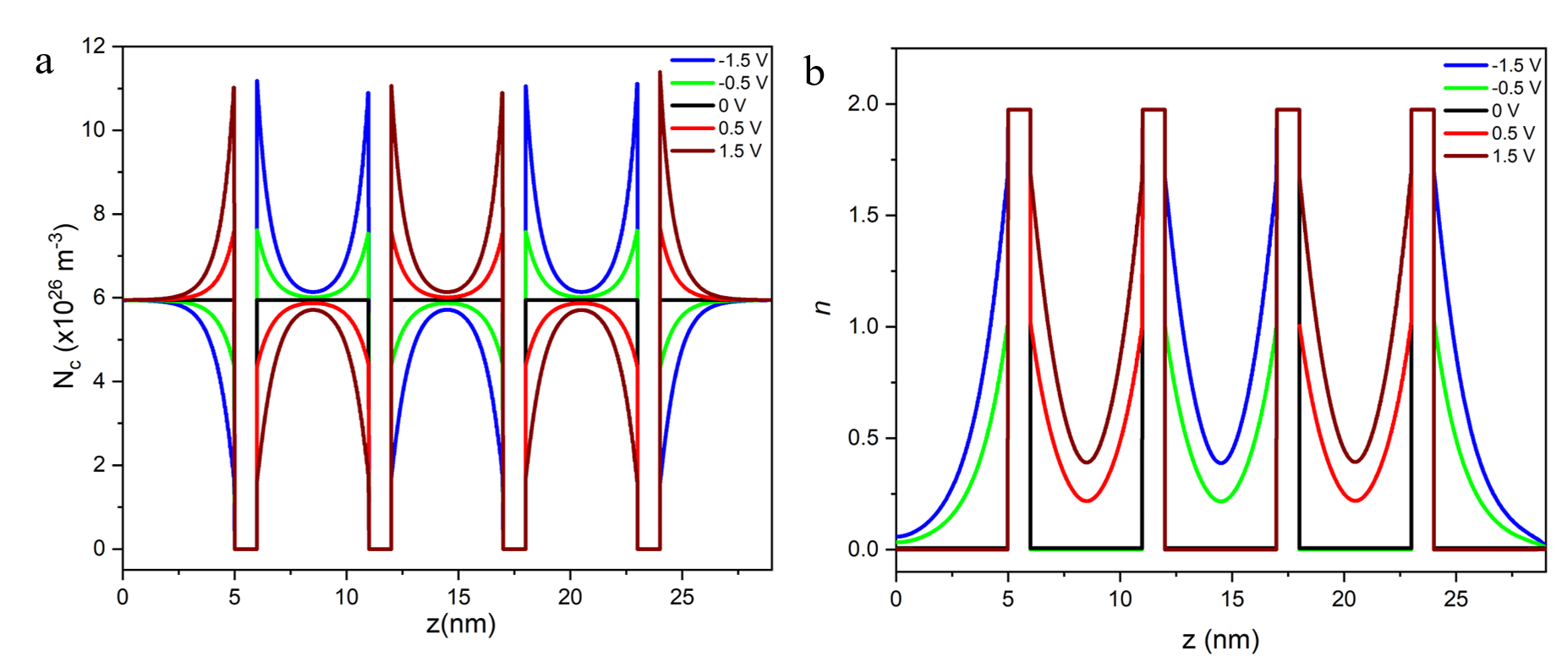}
		\caption{Variation of (a) free carrier density, N$_c$ and (b) real refractive index, $n$ across the length of 5 layer structure as shown in fig. 9b (main paper), for various applied gate voltage $V_G$ } 
		\label{fig:multilayer_nN}
	\end{center}
\end{figure}

\begin{figure}[h]
	\begin{center}
		\includegraphics[width=8cm]{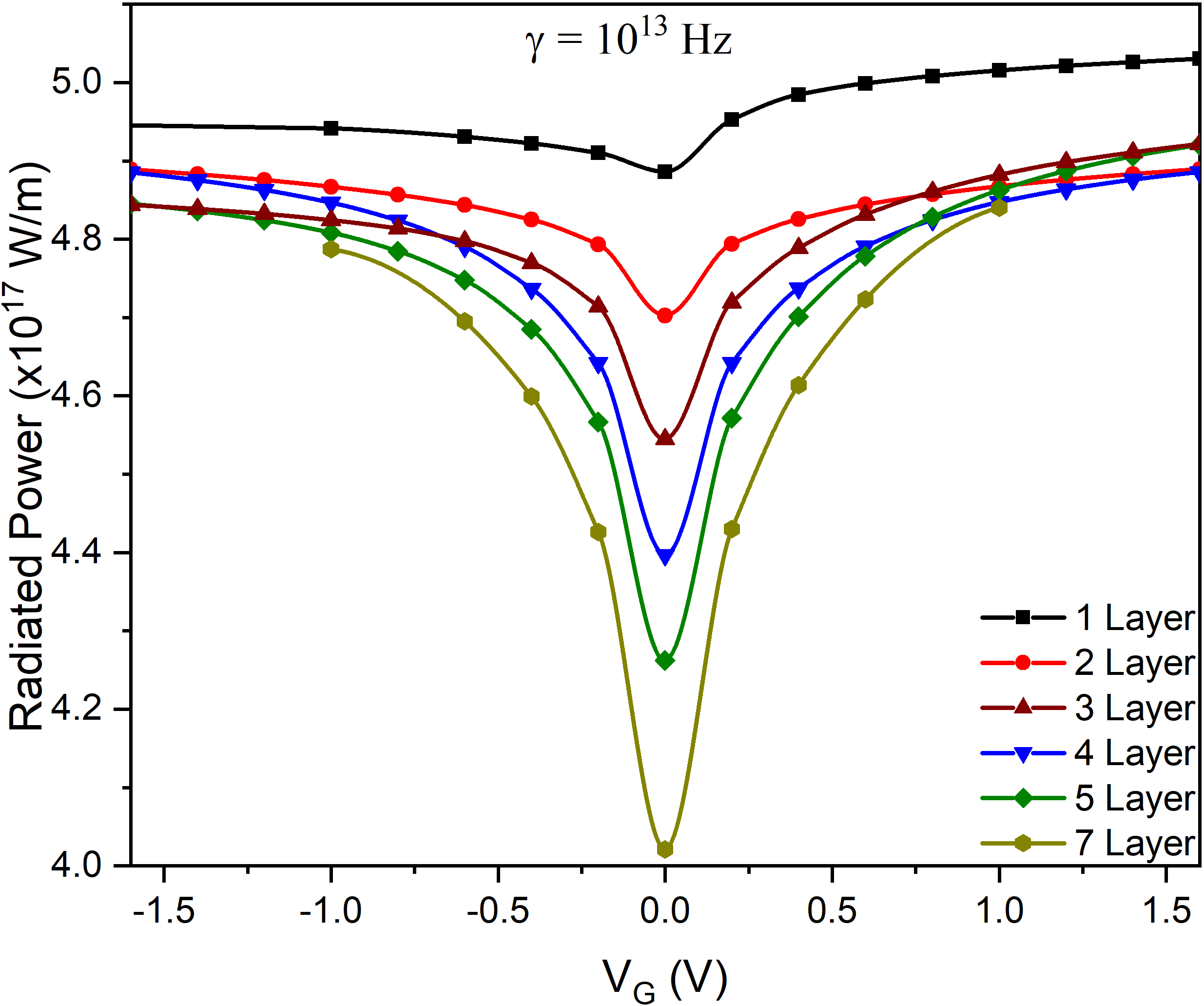}
		\caption{Power radiated as a function of V$_G$ for different number of layers. Here the materials loss is $\epsilon'' = 0.01$}
		\label{fig:multilayer}
	\end{center}
\end{figure}

\subsection{Wider structure for increased angle of steering}

\begin{figure}
	\begin{center}
		\includegraphics[width=15cm]{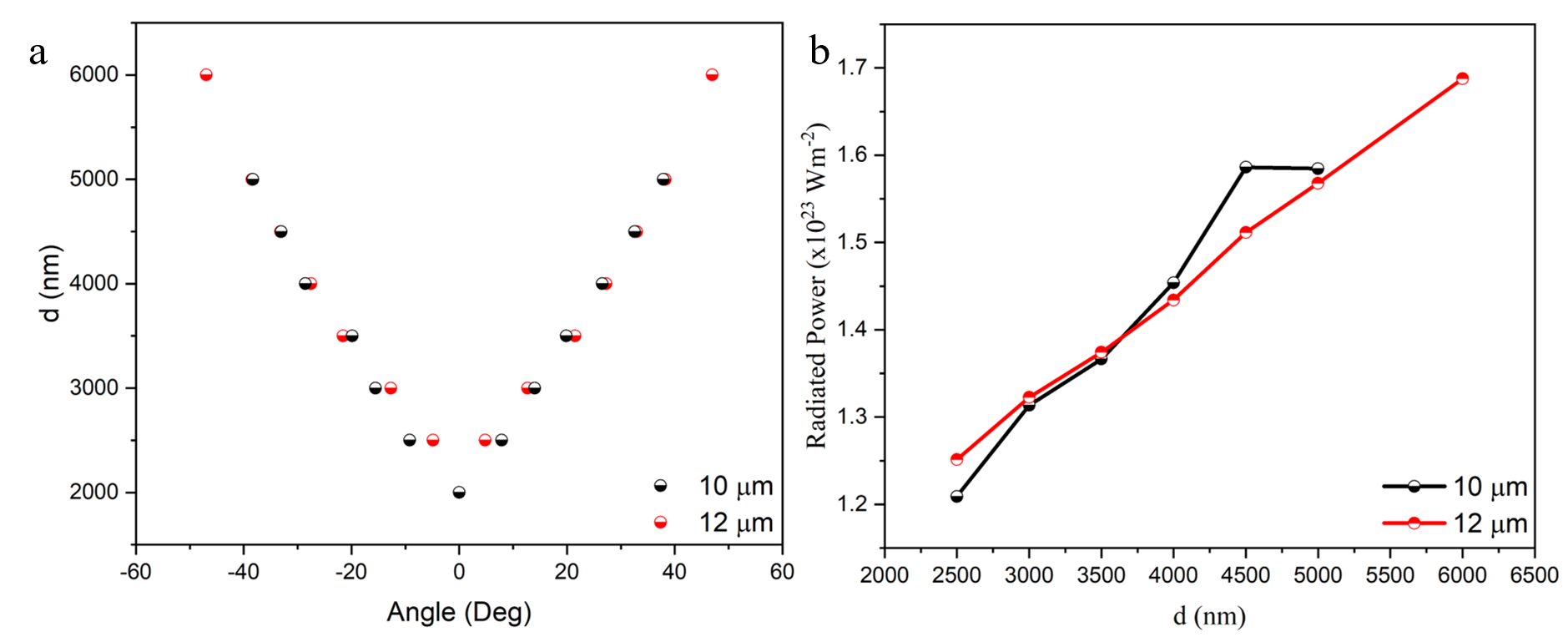}
		\caption{(a) Angle of deviation (b) Maximum power radiated at steered angle as a function of $d$. 10 $\mu$m and 12 $\mu$m denotes the length of the dipole array}
		\label{fig:steering}
	\end{center}
\end{figure}

The simulations demonstrating beam steering were performed with the lateral extent of the dipole array longer than the overall width of the digitated gate electrodes i.e. the maximum $d$. It implies that in real systems the illumination source is always larger than the tunable ENZ strip structures. 
Simulation results presented in fig. \ref{fig:steering} show that the scheme is scalable i.e. the range of steering angle can be increased using wider array of electrodes and a commensurately wider source. The angular deviation increases to $\pm47^\circ$ for the gate electrode array increased to 12 $\mu$m.

\end{document}